%% file: main_v2.tex
\documentclass[twocolumn]{aastex631}

\newcommand{\mbh}{$M_{\rm BH}$}
\newcommand{\mmbh}{M_{\rm BH}}
\usepackage{float}
\usepackage{amsmath}
\usepackage{rotating}

\accepted{March 15, 2024}
\submitjournal{ApJS}

\def\ha{H$\alpha$}
\def\hb{H$\beta$}
\def\mg{Mg{\sc ii}}
\def\civ{C{\sc iv}}
\def\rfe{$R_{\rm FeII}$}

\def\feii{Fe{\sc ii}}
\def\mbh{$M_{\rm BH}$}
\def\lbol{$L_{\rm bol}$}
\def\ledd{$\lambda_{\rm Edd}$}

\def\kms{km s$^{-1}$}

\shorttitle{Virial Black Hole Masses for AGNs behind the Magellanic Clouds}
\shortauthors{Panda et al.}
 
\begin{document}

\title{Virial Black Hole Masses for AGNs behind the Magellanic Clouds}

\author[0000-0002-5854-7426]{Swayamtrupta Panda}\email{spanda@lna.br}\thanks{CNPq fellow}
\affiliation{Laborat{\'o}rio Nacional de Astrof{\'i}sica - MCTI, R. dos Estados Unidos, 154 - Nações, Itajubá - MG, 37504-364, Brazil}

\author[0000-0003-4084-880X]{Szymon Koz{\l}owski}\email{simkoz@astrouw.edu.pl}
\affiliation{Astronomical Observatory, University of Warsaw, Al. Ujazdowskie 4, 00-478 Warsaw, Poland}

\author[0000-0002-1650-1518]{Mariusz Gromadzki}
\affiliation{Astronomical Observatory, University of Warsaw, Al. Ujazdowskie 4, 00-478 Warsaw, Poland}

\author[0000-0002-3051-274X]{Marcin Wrona}
\affiliation{Astronomical Observatory, University of Warsaw, Al. Ujazdowskie 4, 00-478 Warsaw, Poland}

\author[0000-0002-6212-7221]{Patryk Iwanek}
\affiliation{Astronomical Observatory, University of Warsaw, Al. Ujazdowskie 4, 00-478 Warsaw, Poland}

\author[0000-0001-5207-5619]{Andrzej Udalski}
\affiliation{Astronomical Observatory, University of Warsaw, Al. Ujazdowskie 4, 00-478 Warsaw, Poland}

\author[0000-0002-0548-8995]{Michał K. Szyma{\'n}ski}
\affiliation{Astronomical Observatory, University of Warsaw, Al. Ujazdowskie 4, 00-478 Warsaw, Poland}

\author[0000-0002-7777-0842]{Igor Soszy{\'n}ski}
\affiliation{Astronomical Observatory, University of Warsaw, Al. Ujazdowskie 4, 00-478 Warsaw, Poland}

\author[0000-0002-2339-5899]{Pawe{\l} Pietrukowicz}
\affiliation{Astronomical Observatory, University of Warsaw, Al. Ujazdowskie 4, 00-478 Warsaw, Poland}

\author[0000-0001-6364-408X]{Krzysztof Ulaczyk}
\affiliation{Department of Physics, University of Warwick, Coventry CV4 7 AL, UK}
\affiliation{Astronomical Observatory, University of Warsaw, Al. Ujazdowskie 4, 00-478 Warsaw, Poland}

\author[0000-0002-2335-1730]{Jan Skowron}
\affiliation{Astronomical Observatory, University of Warsaw, Al. Ujazdowskie 4, 00-478 Warsaw, Poland}

\author[0000-0002-9245-6368]{Rados{\l}aw Poleski}
\affiliation{Astronomical Observatory, University of Warsaw, Al. Ujazdowskie 4, 00-478 Warsaw, Poland}

\author[0000-0001-7016-1692]{Przemek Mr{\'o}z}
\affiliation{Astronomical Observatory, University of Warsaw, Al. Ujazdowskie 4, 00-478 Warsaw, Poland}

\author[0000-0001-9439-604X]{Dorota M. Skowron}
\affiliation{Astronomical Observatory, University of Warsaw, Al. Ujazdowskie 4, 00-478 Warsaw, Poland}

\author[0000-0002-9326-9329]{Krzysztof Rybicki}
\affiliation{Department of Particle Physics and Astrophysics, Weizmann Institute of Science, Rehovot 76100, Israel}
\affiliation{Astronomical Observatory, University of Warsaw, Al. Ujazdowskie 4, 00-478 Warsaw, Poland}

\author[0000-0002-8911-6581]{Mateusz Mr{\'o}z}
\affiliation{Astronomical Observatory, University of Warsaw, Al. Ujazdowskie 4, 00-478 Warsaw, Poland}

\begin{abstract}
We use the spectroscopic data collected by the Magellanic Quasars Survey (MQS) as well as the photometric $V$- and $I$-band data from the Optical Gravitational Lensing Experiment (OGLE) to measure the physical parameters for active galactic nuclei (AGNs) located behind the Magellanic Clouds. The flux-uncalibrated MQS spectra were obtained with the 4-m Anglo-Australian Telescope and the AAOmega spectroscope ($R=1300$) in a typical $\sim$1.5 hour visit. They span a spectral range of 3700--8500 \AA\ and have S/N ratios in a range of 3--300. We report the discovery and observational properties of 161 AGNs in this footprint, which expands the total number of spectroscopically confirmed AGNs by MQS to 919. After the conversion of the OGLE mean magnitudes to the monochromatic luminosities at 5100 \AA, 3000 \AA, and 1350 \AA, we were able to reliably measure the black hole masses for 165 out of 919 AGNs. The remaining physical parameters we provide are the bolometric luminosities as well as the Eddington ratios. A fraction of these AGNs have been observed by the OGLE survey since 1997 (all of them since 2001), enabling studies of correlations between the variability and physical parameters of these AGNs.
\end{abstract}


\section{Introduction} 
\label{sec:intro}

The black hole mass, \mbh,  in active galactic nuclei (AGNs), is the single most important physical parameter determining most of their properties. It influences the sizes of accretion disks, their innermost stable orbits, temperature profiles, or the spectral energy distribution shapes and luminosities. That is why the black hole mass is the primary parameter sought in AGNs.

Early reverberation mapping campaigns have enabled the first measurements of the black hole masses (e.g., \citealt{1997ASSL..218...85N,2000ApJ...543L...5G,2000ApJ...533..631K}). These campaigns determined simultaneously the distance $R$ to the broad-line-region (BLR) clouds, the time delay $\tau$ between the continuum variability and the responding emission lines ($R=c\tau$), and the velocity $v$ of the BLR clouds. In principle, these two parameters are sufficient to determine the mass, as $\mmbh \propto Rv^2$. \cite{2000ApJ...533..631K} realized that the BLR radius $R$ is tightly correlated with the continuum luminosity $L$, as $R \propto L^{0.7}$, which is known as the radius--luminosity relation for AGNs. The relation was soon improved to yield $R \propto L^{0.5}$ \citep{Bentz_etal_2006, Bentz_etal_2009,Bentz_etal_2013}. Combining the radius-luminosity relation with the equation for the black hole mass, we end up with a simple prescription for the measurement of the black hole mass, as $\mmbh \propto L^{0.5}v^2$. Since both the luminosity $L$ and the velocity $v$ can be simultaneously measured from a single AGN spectrum, it is straightforward nowadays to determine AGN black hole masses for massive spectroscopic surveys with hundreds of thousands of AGN spectra \citep{Shen_etal_2011, Rakshit_etal_2020, Wu_Shen_2022},
albeit with the typical uncertainty of 0.4 dex.

In this paper, we measure the physical parameters (virial black hole masses, luminosities) for AGNs discovered behind the Magellanic Clouds by the Magellanic Quasars Survey (MQS; \citealt{2011ApJS..194...22K,2012ApJ...746...27K,2013ApJ...775...92K}). These two nearby galaxies have been the primary target for microlensing and variability surveys since the early nineties, so a hundred million sources, that can be resolved from Earth, now have two to three decades-long photometric light curves. A combination of the photometric variability and physical parameters for AGNs is a way to improve our understanding of these objects (e.g., \citealt{2009ApJ...698..895K,2016ApJ...826..118K, 2016A&A...585A.129S,2021ApJ...907...96S, Burke_etal_2021Sci}).

In Section~\ref{sec:data}, we present both photometric and spectroscopic data used in our analyses, while in Section~\ref{sec:methods}, we elaborate on the methods used to calculate both monochromatic and bolometric luminosities, 
the methodology of fitting the AGN spectra, and the measurement of basic spectroscopic parameters, in particular FWHM of broad emission lines. This section concludes with the methodology and calculation of the black hole masses for our AGNs along with the corresponding Eddington ratios. The results are presented in Section~\ref{sec:results} and discussed in Section~\ref{sec:discussion}. The paper is summarized in Section~\ref{sec:summary}.

\section{Data} 
\label{sec:data}

In this paper, we analyze spectra for AGNs discovered behind the Magellanic Clouds and obtained by the Magellanic Quasars Survey (\citealt{2011ApJS..194...22K,2012ApJ...746...27K,2013ApJ...775...92K}). The $\sim$4000 spectroscopically observed AGN candidates were selected based on their mid-IR and optical colors, optical variability in the OGLE-III survey, and the X-ray flux. They were observed with the 4-m Anglo-Australian Telescope (AAT) equipped with the AAOmega spectroscope, producing a resolution of $R \approx 1300$ inside a spectral range of 3700--8800 \AA\ in the 580V (blue channel) and 385R (red channel) gratings. Most observations were 1.5h long ($3\times30$ minutes) producing a signal-to-noise ratio of 3--300 with a median of about 40 for $I<19.5$~mag sources (\citealt{2011ApJS..194...22K}). The spectra were reduced with the AAOmega {\sc2dfdr} routines (\citealt{1996ASPC..101..195T}). \cite{2013ApJ...775...92K} reported the discovery of 758 AGNs.

We have re-inspected all the MQS spectra in this analysis and identified 161 additional AGNs, albeit faint (Table \ref{table0-obs}). This makes a sample of analyzed spectra contain a total of 919 AGNs.

We also use the $V$- and $I$-band light curves from the Optical Gravitational Lensing Experiment (OGLE; \citealt{1997AcA....47..319U, 2008AcA....58...69U, 2015AcA....65....1U}) to calculate both the monochromatic and bolometric AGN luminosities. The data have been collected since 1997 with the 1.3-m Warsaw Telescope located in Las Campanas Observatory, Chile. 

A detailed methodology for analyzing both spectra and photometric data is presented in the next section.

\section{Methods} 
\label{sec:methods}

In this section, we provide details of the AGN monochromatic and bolometric luminosities calculation from the OGLE photometry, spectral fitting, and estimating of the black hole masses and Eddington luminosities.

\subsection{Estimating the AGN monochromatic luminosities from photometry}
\label{subsec:lum_estimation_photo}

The primary objective of the Magellanic Quasars Survey was to find and confirm as many AGNs behind the Magellanic Clouds as possible, to measure proper motions of the Clouds (e.g., \citealt{2018ApJ...864...55Z}) and to enable future AGN variability studies.

The spectroscopic AAT observations were taken in sub-optimal weather conditions and without the flux calibration procedure (unnecessary for finding redshifts). Therefore, we are unable to measure the monochromatic (and so bolometric) fluxes directly from these spectra.

\cite{2015AcA....65..251K}, however, provides a method to estimate the monochromatic fluxes from broad-band optical and IR photometry with a typical 0.1 dex uncertainty in a redshift range of $0.1<z<4.9$. Because most AGNs are variable sources, to estimate the weighted mean photometric magnitudes (obtained in the flux space), we use the long-term OGLE data in the $V$- and $I$-band filters, spanning up to 26 years. We then correct the mean observed magnitudes for the extinction using the reddening maps of the LMC and SMC from red clump stars (\citealt{2021ApJS..252...23S}), adopting $A_I=1.217 \times E(V-I)$ and $A_V=2.217 \times E(V-I)$. {Because the \cite{2021ApJS..252...23S} maps provide the median reddening to the red clumps stars, which can be interpreted as the extinction to the center of their distribution in the LMC/SMC, we double the extinction correction to mimic the lines of sight extending all the way though these galaxies.}
We also calculate the $k-$corrections by using the composite SDSS AGN spectrum (\citealt{2001AJ....122..549V}) and OGLE $V$ and $I$ filters (\citealt{2015AcA....65....1U}). Finally, we calculate the
absolute $V$ and $I$ magnitudes for each AGN assuming a standard $\Lambda$CDM cosmological model with ($\Omega_\Lambda$, $\Omega_M$, $\Omega_k$) = (0.72, 0.28, 0.0) 
and $H_0 = 70$ km s$^{-1}$ Mpc$^{-1}$ to calculate the distance modulus (DM). They are provided in Table \ref{table1-obs}.

We then follow the prescription of \cite{2015AcA....65..251K} to calculate ${L_{5100}}$, ${L_{3000}}$, ${L_{1350}}$ monochromatic fluxes (twice, one from the $V$- and one from $I$-band mean magnitude). For each monochromatic flux, we calculate the mean from $V$ and $I$, and the final fluxes are provided in Table \ref{table1-obs}.

\subsection{Spectral fitting with {\sc PyQSOFit}}
\label{subsec:pyqsofit}

We use {\sc PyQSOFit} for spectral decomposition \citep{pyqsofit} of all of our MQS AGN spectra. We correct the spectra to the rest frame and correct for Galactic extinction using the extinction curve of \citet{Cardelli_etal_1989} and the reddening map of \citet{2021ApJS..252...23S}. We then perform a host galaxy decomposition using galaxy eigenspectra from \citet{Yip_etal_2004a} as well as quasar eigenspectra from \citet{Yip_etal_2004b} implemented in {\sc PyQSOFit} code. If more than half of the pixels from the resulting host galaxy fit are negative, then the host galaxy and quasar eigenspectra fit are not applied.

\begin{figure*}[!htb]
    \centering
    \includegraphics[width=\textwidth]{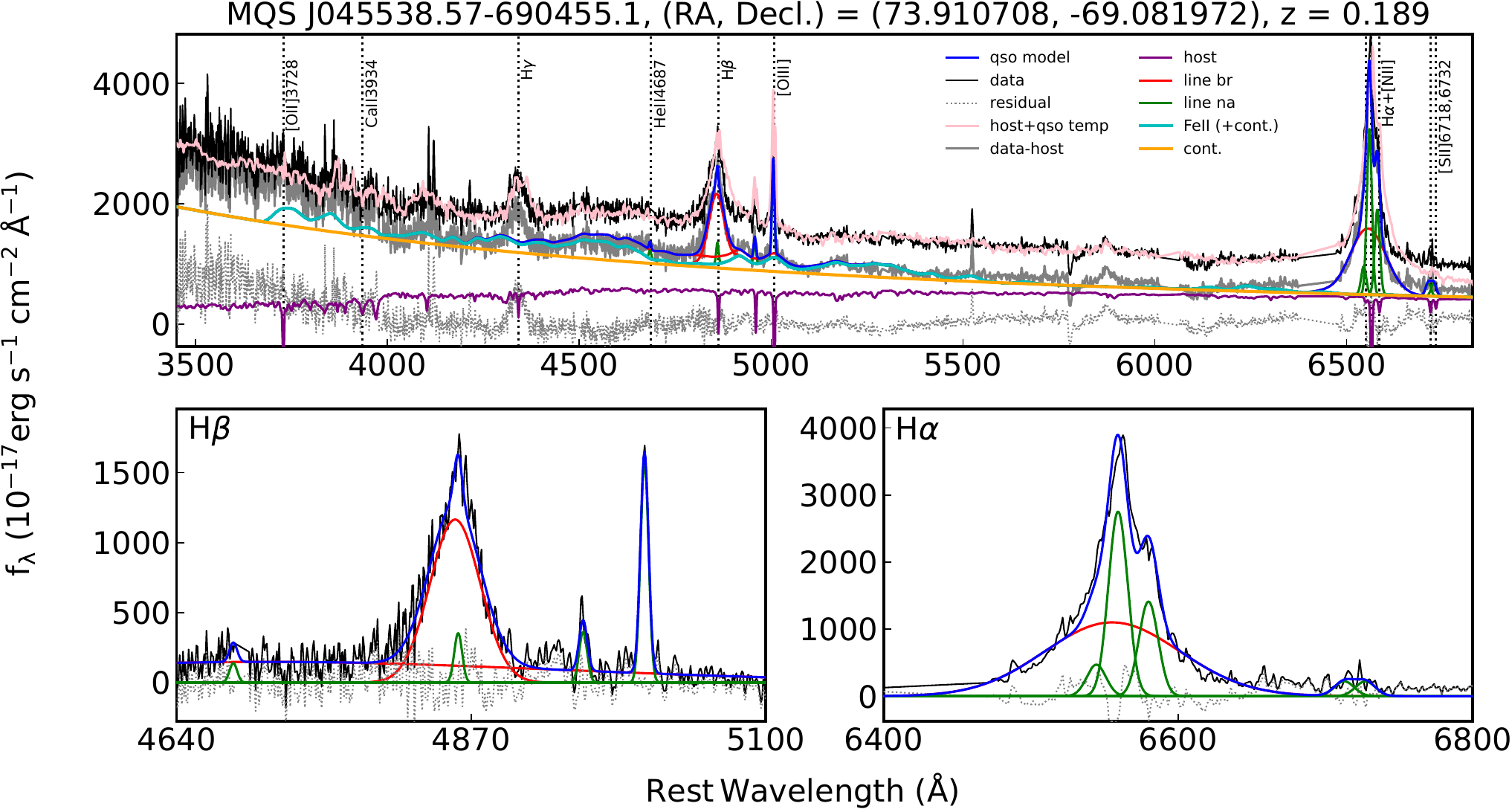}
    \caption{Exemplary fit using {\sc PyQSOFit} \citep{pyqsofit} for a quasar spectrum (MQS J045538.57-690455.1) without significant host-galaxy contribution. In each panel, we show the MQS data (black), power-law continuum (yellow), \feii{} pseudo-continuum ({in addition to the power-law continuum,} light green), broad emission lines (red), narrow emission lines (dark green), the total best-fit {qso} model (blue), which is the sum of continuum and emission lines. {The host galaxy contribution is shown in magenta, while the host subtracted data is shown in a continuous gray line, and the sum of the host and qso model is shown in pink.}  {\it Upper panel}: The rest-frame central wavelengths for prominent emission lines are shown using the dashed vertical lines. The sky coordinates (in degrees), and the redshift for the sources are quoted in the title of the figure. {\it Lower panel}: a zoomed version of individual line complexes. The residuals are shown in {dotted} gray in each panel.}
    \label{fig:pyqsofit-example}
\end{figure*}

We then fit the power law, UV/optical \feii{}, and Balmer continuum models {utilizing the continuum fitting windows as described in \citet{pyqsofit, Rakshit_etal_2020}}. The optical \feii{} emission template spans 3686-7484\AA, from \citet{Boroson_Green_1992}, while the UV \feii{} template spans 1000-3500\AA, adopted from \citet{Vestergaard_Wilkes_2001, Tsuzuki_etal_2006, Salviander_etal_2007}. {\sc PyQSOFit} fits these empirical \feii{} templates using a normalization, broadening, and wavelength shift. Next, we perform emission line fits, using Gaussian profiles as described in \citet{Shen_etal_2019} and \citet{Rakshit_etal_2020}. Depending on redshift and spectral coverage, we fit the following emission lines: H$\alpha\lambda$6564.6 broad and narrow, [NII]$\lambda\lambda$6549,6585, [SII]$\lambda\lambda$6718,6732, H$\beta\lambda$4861 broad and narrow, [OIII]$\lambda\lambda$5007,4959, Mg II$\lambda$2800 broad and narrow, and C IV$\lambda$1549 broad and narrow. In addition, we also fitted the C III]$\lambda$1909 and Ly$\alpha$1216\footnote{We are successful in extracting useful fitted parameters for 13 and 2 sources for C III] and Ly$\alpha$ measurements, respectively.} broad and narrow components but restricted our analysis and results in this work to the sources with fits in the broad H$\alpha$, H$\beta$, Mg II, and C IV broad emission lines. We run all of these fits using Monte Carlo simulation based on the actual observed spectral error array, which in turn yields an error array for all our decomposition fits. An example spectral decomposition is shown in Figure \ref{fig:pyqsofit-example}\footnote{All fitted spectra from our sample can be found on this \url{https://ogle.astrouw.edu.pl/ogle/ogle4/MQS/}.}. The host galaxy fits used in {\sc PyQSOFit} are limited to rest-frame wavelengths between 3450 – 8000 \AA. Due to this limitation, to fit the \mg{} line complex, we follow the prescription of \citet{Green_etal_2022} but make a conditional execution of host decomposition in the same run, i.e., if $z$ $<$ 0.25, then the host contribution is included. Otherwise, the host contribution is not accounted for. {In our fitting routine using PyQSOFit we fit the spectrum over the whole wavelength range, although for this work, we only use the measurements of the FWHMs of the broad emission lines (\hb{}, \mg{}, and \civ{}). These profiles are fitted within narrow wavelength windows ($\sim$100-150\AA, e.g., as shown in the bottom panels in Figure \ref{fig:pyqsofit-example}) after the power-law continuum and host contribution are removed leaving only the emission line profiles to be fitted, and the effect from the absolute spectrophotometric calibration is minimal. This primarily affects the estimation of the continuum luminosities and therefore, we make use of the OGLE photometry-derived monochromatic luminosities throughout this work.} We report the FWHMs for the \hb{}, \mg{}, and \civ{} emission lines for our sources in Table \ref{table2-obs}.

\subsection{Estimating black hole mass and Eddington ratios}
\label{subsec:bh_mass}

To calculate the bolometric luminosity (\lbol{}), we follow the prescription of \citet{Richards_etal_2006, Shen_etal_2011, Rakshit_etal_2020}, where the AGN monochromatic luminosity is scaled by a bolometric correction factor to estimate the \lbol{}:

\[
    {L_{bol}}= 
\begin{cases}
    9.26\times {L_{5100}} & \text{if } z<0.8\\
    5.15\times {L_{3000}} & \text{if } 0.8\leq z<1.9\\
    3.81\times {L_{1350}} & \text{if } z\geq 1.9\\
\end{cases}
\]

Next, the black hole mass (\mbh{}) can be estimated using the virial relation from the single-epoch spectrum for which continuum monochromatic luminosity (here derived from photometry) and line width measurements are available using the following relation:

\begin{equation}
    \log \left(\frac{{M_{BH}}}{{\rm M_{\odot}}}\right) = A + B\log \left(\frac{{L_{\lambda}}}{10^{44} {\rm \;erg\;s^{-1}}}\right) + 2\log \left(\frac{{\rm FWHM}}{{\rm km\;s^{-1}}}\right),
    \label{eq:mbh}
\end{equation}

where A and B are the constants empirically calibrated from prior studies. Following the prescription of \citet{Rakshit_etal_2020}, we used the black hole mass calibrations from \citet[][hereafter VP06]{Vestergaard_Peterson_2006}, and \citet[][hereafter VO09]{Vestergaard_Osmer_2009}:

\[
    A, B = 
\begin{cases}
    (0.910, 0.50) & \text{for } {\rm H\beta}, \text{VP06}\\
    (0.860, 0.50) & \text{for } {\rm MgII}, \text{VO09}\\
    (0.660, 0.53) & \text{for } {\rm CIV}, \text{VP06}\\    
\end{cases}
\]

Subsequently, we estimate the Eddington ratio (\ledd{}), i.e. the ratio of the \lbol{} to the Eddington luminosity\footnote{$L_{\rm Edd} \approx 1.26\times 10^{38} \left(\frac{M_{BH}}{\rm M_\odot}\right)$ erg s$^{-1}$.} (L$_{\rm Edd}$).  The derived \lbol{}, \mbh{}, and \ledd{} for the sources in our sample are reported in Table \ref{table3-obs}. We do not account for the error on the constant term (A) while estimating the uncertainties on the BH masses (Sec.~\ref{subsec:bh_mass_errors}).

\subsection{Error budget}
\label{subsec:bh_mass_errors}
In this subsection, we discuss the error budget for the black hole mass and luminosity measurements.

The black hole mass equation (Equation~\ref{eq:mbh}) contains four variables ($A$, $B$, ${L_{\lambda}}$, and FWHM). While the uncertainties for $A$ and $B$ typically are not reported, the usual dispersion of this relation is 0.3--0.4 dex (e.g., VP06). This means that a single measurement of the black hole mass has an uncertainty of about 0.4 dex. 

The black hole mass, via Equation~(\ref{eq:mbh}), also depends on FWHM, which is estimated along with its uncertainty by PyQSOfit from fitting the broad emission lines, and the monochromatic luminosity. The uncertainty of the latter is estimated as a combination of two factors: (1) the uncertainty of the conversion of the broad-band magnitudes to the monochromatic luminosities which is typical of the order of 0.1 dex, and (2) the uncertainty of the mean broad-band magnitude. The latter one depends not only on the photometric quality of the survey but also on the data length and number of points as AGNs are variable sources. The longer the light curve and the larger the number of photometric points, the closer the mean estimation is to the true mean. The contribution of this uncertainty to the total uncertainty is 0.004 dex.

\section{Results} 
\label{sec:results}

\subsection{Luminosity distribution as a function of  redshift}
\label{subsec:lum_dist}

\begin{figure*}[htb!]
    \centering
    \includegraphics[width=\textwidth]{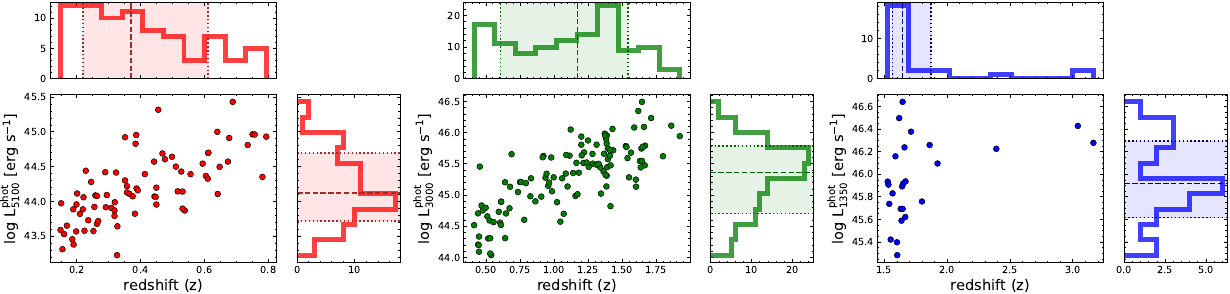}
    \caption{AGN luminosity at 5100\AA~({\it left panel}), 3000\AA~({\it middle panel}), and 1350\AA~({\it right panel}) derived from the OGLE photometry as a function of redshift. The respective marginal distributions are shown per panel. The medians are marked with dashed lines and the shaded regions mark the region between 16$^{\rm th}$ and 84$^{\rm th}$ percentiles of the distributions in each of the marginal distributions.}
    \label{fig:L_v_z}
\end{figure*}

In Figure \ref{fig:L_v_z}, we demonstrate the dependence of the monochromatic luminosity on the redshift for the sources in our sample. The luminosities are derived from the photometry as described in Sec. \ref{subsec:lum_estimation_photo}. We highlight three cases of monochromatic luminosities: (a) at 5100\AA, (b) at 3000 \AA, and (c) at 1350 \AA, which are in the vicinity of the prominent broad emission lines, i.e., \hb{}, \mg{}, and \civ{}, respectively. The properties (median and the 16th and 84th percentiles) for the respective joint distributions presented in Figure \ref{fig:L_v_z} are reported in Table \ref{tab:table_lum_vs_z}. 

\input{table_lum_vs_z}

\begin{figure}[htb!]
    \centering
    \includegraphics[width=\columnwidth]{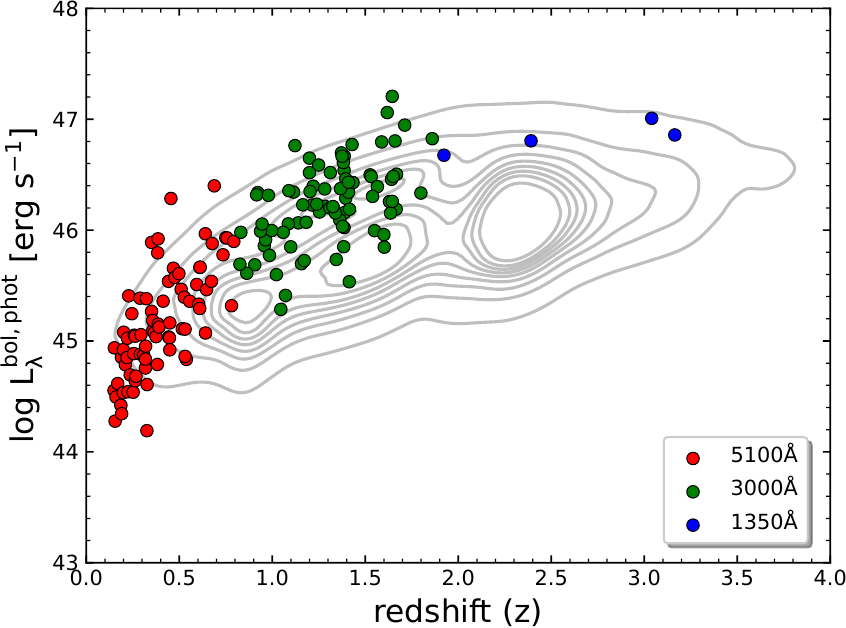}
    \caption{Bolometric luminosity as a function of redshift for the sources in our sample. The sources are colored by the AGN monochromatic luminosity used to estimate the bolometric luminosity. The sources from the SDSS DR14 QSO sample are shown using contours where the fiducial \lbol{} are reported with the {\sc qual} flag = 0. We show 9 levels for the contour map, which correspond to the isodensity of the SDSS sources, with the outermost contour encompassing 90\% of the sources and decreasing inwards by  10\%.}
    \label{fig:Lbol_v_z}
\end{figure}

To facilitate the comparison of the sources and their luminosities across redshift, we estimate the bolometric luminosities (using the prescription outlined in Sec. \ref{subsec:bh_mass}). Figure \ref{fig:Lbol_v_z} demonstrates the bolometric luminosity (\lbol{}) as a function of redshift for all of the sources in our sample. The sources are colored based on the monochromatic luminosity used to estimate the respective \lbol{} values. We see a clear increase in the net \lbol{} with increasing redshift extending up to $z$ $\sim$ 3.5, where the bottom envelope is due to the limiting magnitude of the SDSS or MQS/OGLE surveys. %{We, however, note the saturation in \lbol{} (i.e. $\sim$ 47, in log-scale) for sources above z$\gtrsim$1.6, which are limited in number. These sources which are at the highest redshifts, owing to the estimation of the \lbol{} using the 1350\AA\, show a slight offset to those estimated using the 3000\AA{}.}

We overlay a filtered version of the SDSS DR14 QSO sample \citep{Rakshit_etal_2020} on this distribution to compare the two distributions. The filtering of the sources is made by using the quality flags associated with the \mbh{} and \lbol{} estimations from \citet{Rakshit_etal_2020}. They use {\sc QUAL} flag = 0 to identify sources with reliable \mbh{} and \lbol{} estimations. We have chosen to use the two flags simultaneously to avoid confusion later when discussing the \mbh{} measurements from SDSS and our sample. The original SDSS DR14 QSO sample contains 526,265 sources of which, after filtering, we are left with 449,863. We note here that before filtering the sample of sources in the SDSS many erroneous estimates were reported for the \lbol{} and \mbh{} with significantly large uncertainties. The filtering allowed us to remove sources with such measurements and limit ourselves to estimates with higher reliability. To highlight the large differences due to the filtering, we report the median, minimum, and maximum values for the redshift, \lbol{} and \mbh{} distributions for the original and the filtered SDSS samples in Table \ref{tab:table_SDSS}. In Figure \ref{fig:Lbol_v_z}, we show the filtered SDSS sample using contours. We use 9 levels for the contour map which correspond to the iso-density lines encompassing 90\% of the SDSS AGNs (the outermost contour) and decreasing inwards by 10\%. We note that some of the sources in our sample (28) lie outside the lowest contour line, which is the consequence of differences in the surveys' setups. We see an overall agreement between the two distributions with a clear increase in the \lbol{} with increasing redshift. Additionally, we note that sources in our sample have relatively higher \lbol{} values as compared to the peak of the SDSS distribution irrespective of the monochromatic luminosity used to estimate the \lbol{} values. This can be attributed to the shallower depth of the OGLE survey as compared to the SDSS survey.

\input{table_SDSS}

\subsection{Black hole mass and Eddington ratio distributions}
\label{subsec:mbh_edd_dist}

\begin{figure*}[htb!]
    \centering
    \includegraphics[width=0.32\textwidth, height=0.2475\textwidth]{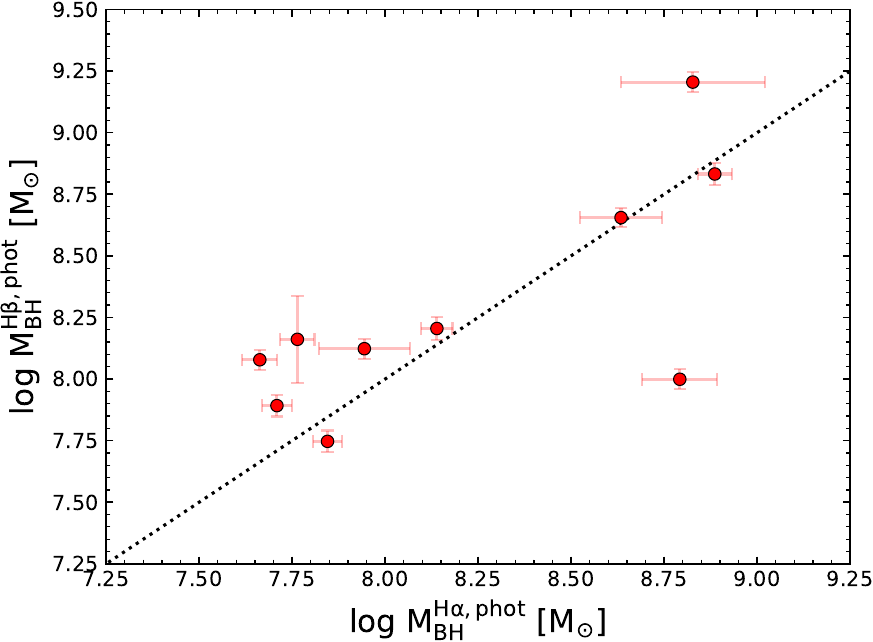}
    \includegraphics[width=0.32\textwidth]{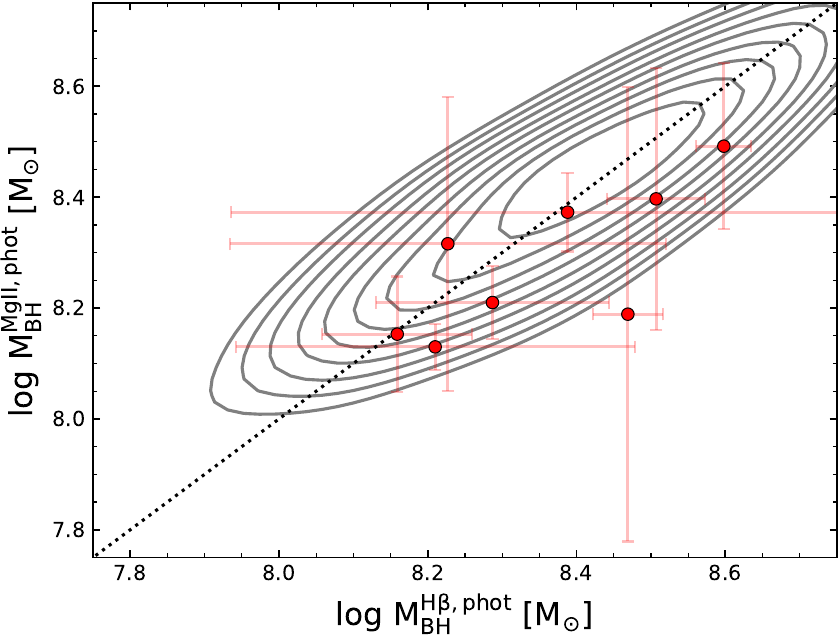}
    \includegraphics[width=0.32\textwidth]{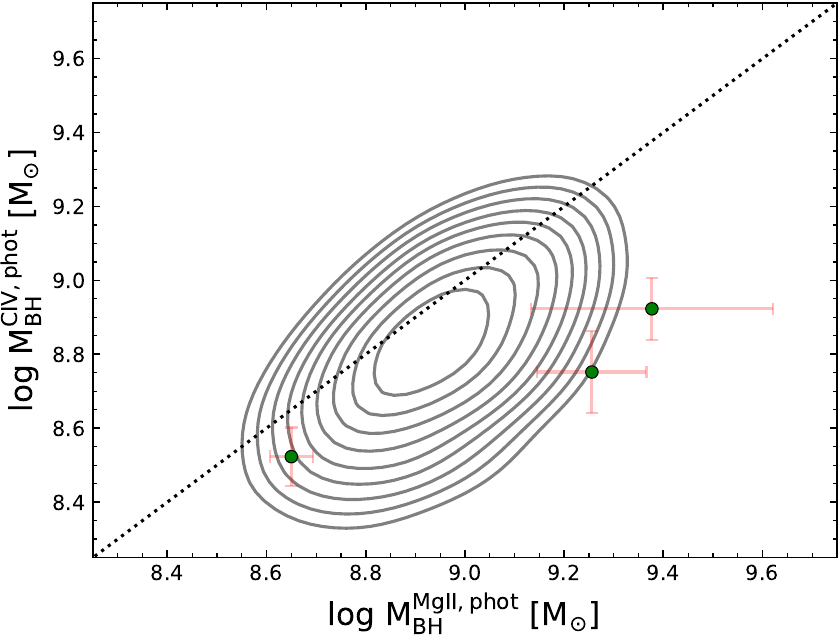}        
    \caption{Comparison of the black hole mass estimates, {\it left panel:} for sources where both \ha{} and \hb{} FWHMs are simultaneously available; {\it middle panel:} for sources where both \hb{} and \mg{} FWHMs are simultaneously available; and {\it right panel:} for sources where both \mg{} and \civ{} are simultaneously available in a spectrum. In each panel, the {dotted} black line represents the 1-to-1 line shown for reference. The sources from the SDSS DR14 QSO sample are shown using contours where the \hb{}-, \mg{}- and \civ{}-based BH masses are reported (no \ha{}). For the SDSS sample, we filter the sources based on the fiducial \lbol{} and \mbh{} with the {\sc qual} flags = 0. The largest contour represents 67\% of the total number of the SDSS AGNs.}
    \label{fig:mbh_compare}
\end{figure*}

In Figure \ref{fig:mbh_compare}, we demonstrate the \mbh{}-\mbh{} planes estimated using the pairs of emission lines, i.e., (\ha{}, \hb{}), (\hb{}, \mg{}) and (\mg{}, \civ{}), respectively. We have 10 sources with simultaneously reliable \ha{}-based and \hb{}-based \mbh{} measurements in our sample. Similarly, we have eight sources with reliable \hb{}-based and \mg{}-based \mbh{} measurements, and three sources with reliable \mg{}-based and \civ{}-based \mbh{} measurements. Overall, we find a good agreement between the masses estimated using FWHM from different emission lines and monochromatic luminosities, which are depicted using the line of unity (dashed line) in each panel of Figure \ref{fig:mbh_compare}. The scatter in the panels of Figure \ref{fig:mbh_compare} can be attributed to either the relative offsets in the FWHM values between the lines, the monochromatic luminosities differences, and the uncertainty in the relations (i.e., mostly due to the constant term (B) associated with the monochromatic luminosity) used to derive the \mbh{}. We note, however, that we do not account for the error on the constant term (A, see Equation~\ref{eq:mbh}) while estimating the uncertainties on the BH masses.

Similarly to Figure \ref{fig:Lbol_v_z}, in Figure~\ref{fig:mbh_compare} we overlay the contours from the filtered SDSS sample. The SDSS catalog provides the \mbh{} mass measurements obtained using the \hb{}, \mg{}, and \civ{} emission lines and respective monochromatic continuum luminosities (no \ha{}). Hence, we only show these contour maps for the middle (\hb{}-based \mbh{} versus \mg{}-based \mbh{}) and right (\mg{}-based \mbh{} versus \civ{}-based \mbh{}) panels. Contrary to the contour maps in Figure \ref{fig:Lbol_v_z}, we truncate the contours at 67\% and above for the probability mass for the respective distributions for better visualization of the comparison between the two samples. We notice that all the measurements from our sample, including the uncertainties, lie within the threshold of the filtered SDSS sample.

\begin{figure}[htb!]
    \centering
    \includegraphics[width=\columnwidth]{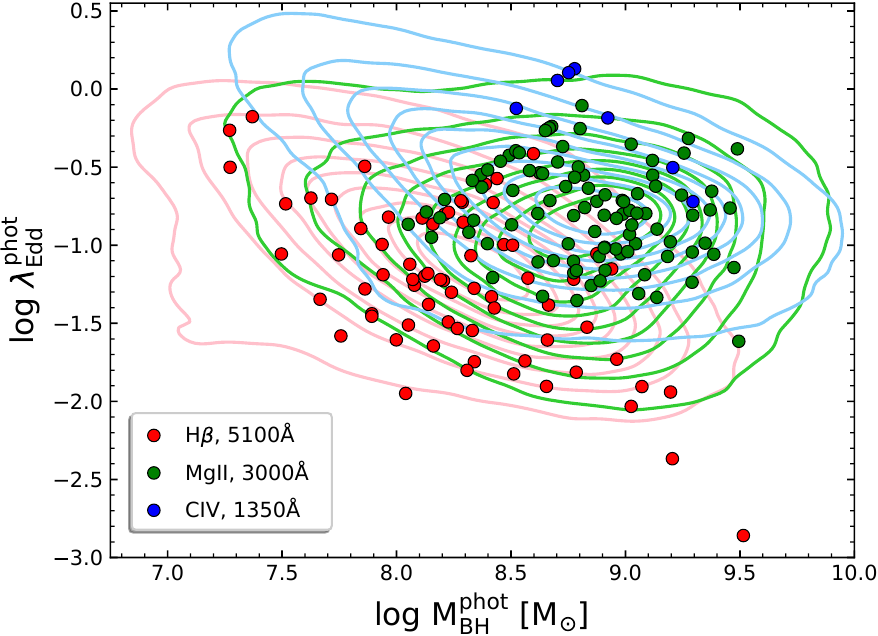}
    \caption{Black hole masses versus Eddington ratios for the sources in our sample. The sources are colored based on the monochromatic luminosity-emission line pairs. The sources from the SDSS DR14 QSO sample are shown using contours where the \hb{}- (in pink), \mg{}- (in light green), and \civ{}-based (in light blue) BH masses are considered. The SDSS sources are filtered based on the adopted fiducial \lbol{} and \mbh{} with the {\sc qual} flags = 0. {We do not show the uncertainties associated with the \mbh{} and \ledd{} for the MQS AGNs for clarity.}}
    \label{fig:Mbh_v_Eddr}
\end{figure}

\input{table_Eddr_MBH_MQS166}

\begin{figure}
    \centering
    \includegraphics[width=\columnwidth]{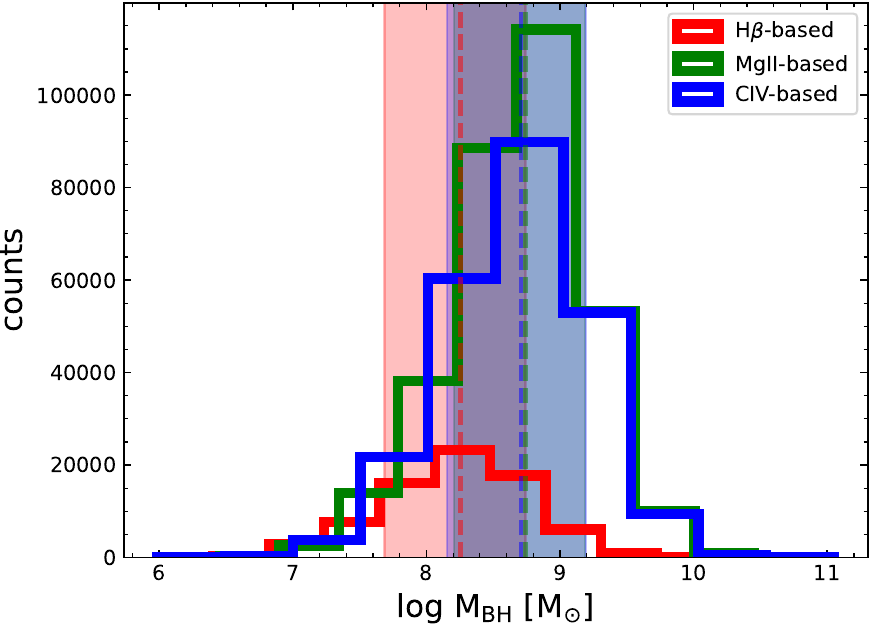}
    \caption{{\mbh{} distributions from the \textit{cleaned} SDSS DR14 QSO sample \citet{Rakshit_etal_2020}. The \hb{}- (red), \mg{}- (green), and \civ{}-based (blue) masses are shown in this histogram. The respective median values are shown using dashed lines of identical colors, while the shaded regions mark regions bounded by the 16$^{\rm th}$ and 84$^{\rm th}$} percentiles for the respective distributions.}
    \label{fig:mbh-dist}
\end{figure}

In Figure \ref{fig:Mbh_v_Eddr}, we present the Eddington ratio (\ledd{})--\mbh{} plane occupied by the sources in our sample. The sources are colored based on the respective emission lines and monochromatic continuum luminosities incorporated to estimate the \ledd{} and \mbh{}. We notice that the distribution shifts towards higher black hole masses and higher Eddington ratios as we move from (\hb{}, 5100\AA) based sub-sample to (\mg{}, 3000\AA) and (\civ{}, 1350\AA) sub-samples. This trend is also quantified in Table \ref{tab:table_Eddr_MBH}, where we see that the ranges covered by the \hb{}-based \mbh{} and \ledd{} are the widest. While the \mg{}-based sample is more concentrated at a slightly larger \mbh{} range but covers a subset of the range in the Eddington ratio relative to the \hb{}-based sub-sample. Finally, the \civ{}-based sample only contains seven sources, much smaller than the other two sub-samples (we have 70 and 97 sources, for the \hb{}-based and \mg{}-based sub-samples, respectively.), and occupies a region with the highest \ledd{} even going above the Eddington limit. Although, the \mbh{} range is relatively modest as compared to the other two sub-samples. Similar to the previous analyses, we overlay the corresponding contour maps from the filtered SDSS sample for the respective sub-samples. In this figure, the contour maps show the full range of the distribution from the filtered SDSS sample without any threshold cuts. We notice that the sources from both our and the SDSS sample occupy roughly the same region in the \ledd{}-\mbh{} plane, albeit a few sources from the \hb{} sub-sample from our sample which has slightly lower Eddington ratio values as compared to their SDSS counterparts. {We note here in passing, that the masses derived using the \civ{} region and \mg{} region are comparable for our MQS quasars sample. This similarity between the \mbh{} estimates is also noted in the \mbh{} distributions derived from the SDSS DR14 sample (see Figure \ref{fig:mbh-dist}). We consider the sources where the quality flags for the \mbh{} is 0, i.e., the masses measurements are reliable. We independently show the masses estimated from the \hb{}, \mg{}, and \civ{} regions which use the formalisms from \citet{Vestergaard_Peterson_2006}, \citet{Vestergaard_Osmer_2009}, and \citet{Vestergaard_Peterson_2006}, respectively. The median values for each sub-sample are shown using dashed lines and the region between the 16$^{\rm th}$ and 84$^{\rm th}$ percentiles are shown using shaded colors per sub-sample. For \mg{}- (green) and \civ{}-based (blue) sub-samples, we find that the distributions behave similarly, i.e., the respective medians are comparable (8.74 vs 8.71) and the regions bounded by the 16$^{\rm th}$ and 84$^{\rm th}$ percentiles also overlap. The overall similarity in \mbh{} using different broad emission lines has been noted in other studies \citep{2011ApJ...742...93A, 2017ApJS..228....9K}. We, however, note that the mass measurements can be affected by the choice of methodology \citep[see e.g.,][]{Mejia-Restrepo_etal_2018, 2020ApJ...903..112D}.}

\subsection{Optical plane of the Eigenvector 1}
\label{subsec:qms}

\begin{figure*}[htb!]
    \centering
    \includegraphics[width=0.425\textwidth]{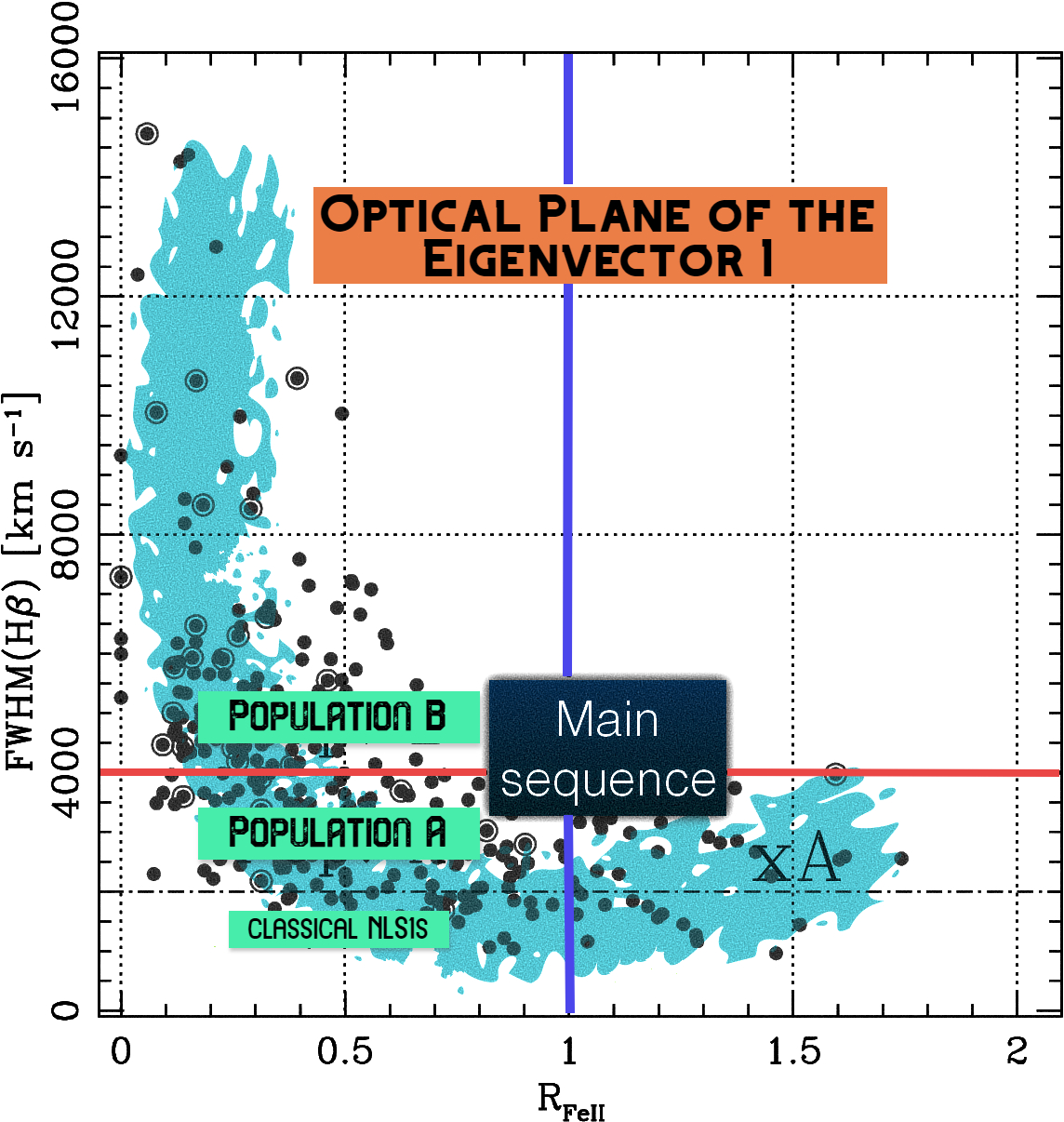}
    \includegraphics[width=0.565\textwidth, height=0.4675\textwidth]{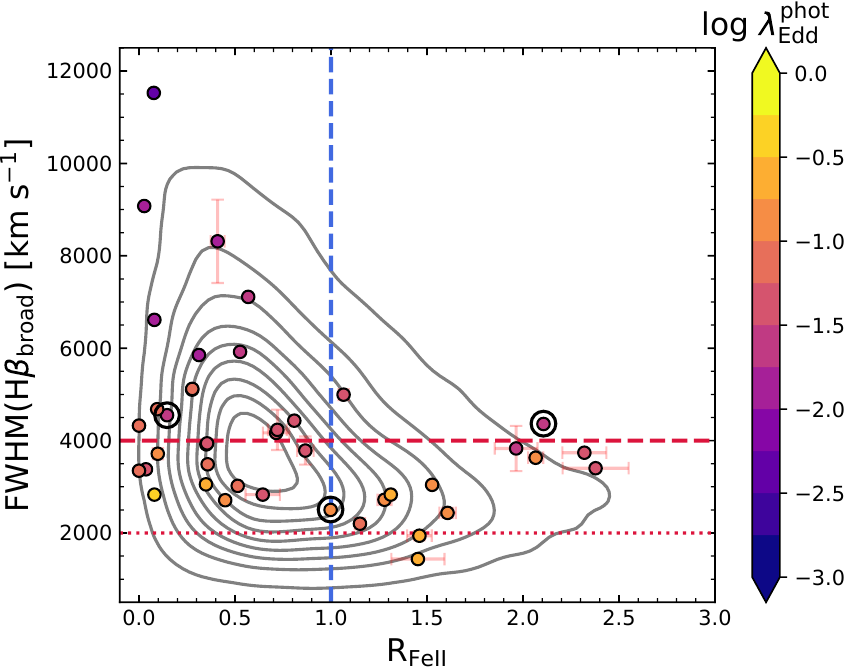}
    \caption{{\it Left panel}: Schematic diagram of the optical plane of the Eigenvector 1. Abridged version from \citet{panda19c}. The horizontal line denotes the threshold in FWHM(\hb{}) at 4000 \kms{} which separates the Population A and Population B sources. The ``classical'' NLS1s are located below the FWHM(\hb{}) $\leq$ 2000 \kms{} (dot-dashed line). The vertical blue line marks the limit for \rfe{} = 1 separating the weak and strong \feii{} emitters (or xA sources). {\it Right panel}: The optical plane for the MQS sources. Similar to the left panel, the horizontal dashed and dotted lines represent the 4000 \kms{} and 2000 \kms{} thresholds, respectively, while the dashed vertical line marks the \rfe{} = 1 limit. The sources are colored based on their Eddington ratios (in the log scale). Here, we demonstrate the sources where both the FWHM(\hb{}) and \rfe{} are of high quality, i.e., corresponding errors are within 20\% of the estimated values. The sources from the SDSS DR14 QSO sample are shown using contours where a similar quality filtering is adopted. Spectra for the three sources marked with the bulls-eyes are shown in Figure \ref{fig:compare_spectra}.}
    \label{fig:EV1}
\end{figure*}

Understanding the diversity in spectral properties within AGNs poses a significant challenge. To this end, the work by \citet{Boroson_Green_1992} holds paramount importance for two key reasons. Firstly, it represents one of the pioneering contributions in AGN research employing Principal Component Analysis (PCA) to unravel the interrelation between observed quasar properties. This analysis delves into the Main Sequence of Quasars, employing Eigenvectors, notably Eigenvector 1. This particular eigenvector reveals an intriguing anti-correlation between the equivalent width (EW) of the optical \feii{} blend (spanning 4434-4684 Å) and the peak intensity of the forbidden line [OIII]$\lambda$5007 Å. Secondly, this study also establishes a connection between the FWHM of the broad H$\beta$ emission and this eigenvector. This linkage, specifically between the FWHM of the broad H$\beta$ line and the strength of the \feii{} blend (expressed as EW(\feii{}) relative to the EW of the broad component of H$\beta$, or \rfe{}), has evolved into the well-established ``Quasar Main Sequence''. This sequence, illustrated in the left panel of Figure \ref{fig:EV1}, is primarily influenced by the Eddington ratio among other physical properties, as documented in subsequent studies \citep[e.g.,][]{sul00,sh14,mar18,panda18b,panda19,panda19b}.

\input{table_QMS}

Furthermore, an additional classification system based on the width of the H$\beta$ emission line profile in AGN spectra has been introduced, distinguishing between Population A and Population B. Population A encompasses local Narrow-Line Seyfert 1 galaxies (NLS1s) and more massive high accretors, primarily identified as radio-quiet sources \citep[e.g.,][]{ms14}, with FWHM(H$\beta$) $\lesssim$ 4000 km s$^{-1}$. Notably, Population A sources often exhibit a H$\beta$ profile with a Lorentzian-like shape \citep[e.g.,][]{sul02,zamfiretal10}. In contrast, Population B sources, characterized by broader H$\beta$ profiles ($\gtrsim$ 4000 km s$^{-1}$), are predominantly associated with ``jetted'' characteristics \citep[e.g.,][]{padovani2017}. These sources tend to exhibit Gaussian-shaped H$\beta$ profiles, and for those with even higher FWHMs, disk-like double Gaussian profiles are observed in Balmer lines. The choice of the FWHM cutoff at 4000 km s$^{-1}$ was proposed by \citet{sul00,mar18}, who observed more pronounced changes in AGN properties beyond this line-width threshold. Subsequent studies have shown that the two populations form a continuous link and share a connection \citep{fraix-burnet2017,berton2020}. The morphology of the emission line profiles and the characteristics of the continuum are intricately linked to the central engine, specifically the black hole mass, accretion rate, black hole spin, and the viewing angle from a distant observer \citep{czerny2017, mar18, panda18b, panda19b, Panda_PhD}.

In our sample, to check the location of the sources on the Eigenvector 1 sequence, we first filtered the sources where the relative uncertainties on the \rfe{} and the FWHM(\hb{}) were below a certain threshold. We assume this limit to be 20\% to keep reasonable measurements and avoid sources where these values could be unreliable or affected by low signal-to-noise spectral quality. This limits the total number of sources to 41/58, where 58 was the source count where we have a non-zero measurement for the \rfe{} and FWHM(\hb{}). We tabulate the salient properties of this limited sample of 41 sources in Table \ref{tab:table_qms}. In the right panel of Figure \ref{fig:EV1}, we demonstrate the optical plane of the Eigenvector 1 sequence for our sources. These sources are color-coded by their respective \ledd{} values. To facilitate the comparison between our sample and the filtered SDSS sample, we overlay the SDSS sample using contour maps. We find remarkable agreement between the two samples. Some of the sources from our sample do have slightly larger FWHM(\hb{}) and/or larger \rfe{} estimates. We note, however, that the exact extent of the filtered and limited SDSS sample considered here does have a wider coverage (please see the lower half of the Table \ref{tab:table_qms}), although these sources constitute a minor fraction of the total sample considered here.

\begin{figure}[htb!]
    \centering
    \includegraphics[width=\columnwidth]{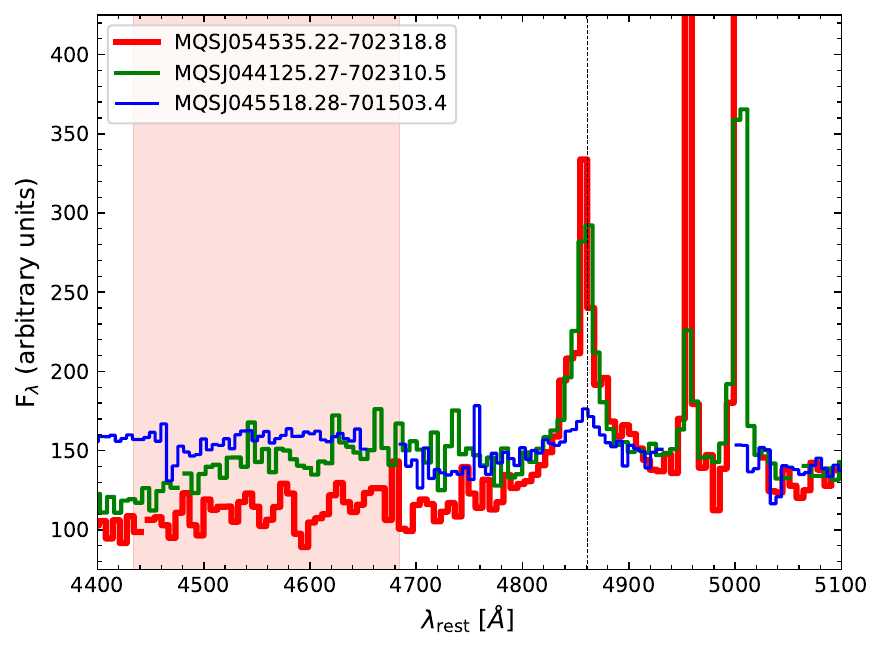}
    \caption{Binned spectra for the sources highlighted in the right panel of Figure \ref{fig:EV1}, corresponding to three \rfe{} regimes - low (in red, 0.146$\pm$0.005), medium (in green, 0.998$\pm$0.009), and high (in blue, 2.1068$\pm$0.018). The shaded region shows the \feii{} range (4434--4684\AA) and the central wavelength for the \hb{} is marked using the dotted line.}
    \label{fig:compare_spectra}
\end{figure}

We highlight the spectra of three of our sources in Figure \ref{fig:compare_spectra}. These three sources are marked with bullseye symbols in Figure \ref{fig:EV1} and were chosen to demonstrate the variety in \rfe{} measurements we have in our sample. The spectra have been binned for better visualization. We can see that going from the source with one of the lowest \rfe{} measurements (spectrum in red), towards the source with one of the highest values for \rfe{} (spectrum in blue), we see a substantial change in the \feii{} bump feature and the weakening of the \hb{} emission. This further demonstrates the efficacy of the quasar main sequence analysis and its potential to categorize a diverse population of Type-1 AGNs.

\section{Discussion} 
\label{sec:discussion}

Measuring the physical parameters of AGNs appears to be a straightforward task nowadays, as a single spectrum for an AGN is typically necessary to measure the monochromatic and bolometric luminosities, the black hole mass, and the Eddington ratio.  The prime example, where such measurements were reported for 526,265 AGNs is the SDSS DR14 QSO catalog by \citet{Rakshit_etal_2020} and the SDSS DR16 version with 750,414 AGNs (\citealt{Wu_Shen_2022}).

While the black hole mass sets the size of an accretion disk, the key to our understanding of the physical processes within the disk may be ciphered in the observed AGN variability patterns. Several theoretical variability timescales are predicted: the dynamical one, which is the time it takes the matter to orbit the black hole ($\tau_{\rm dyn}=\sqrt{GM/r^3}$), the thermal timescale ($\tau_{\rm th}=\alpha^{-1}\tau_{\rm dyn}$), or the viscous timescale ($\tau_{\rm vis}=\tau_{\rm th}(r/h)^2$), where $M$ is the black hole mass, $r$ is the radial size of the disk, $\alpha$ is the viscosity, and $h$ is the disk height (e.g., \citealt{2004astro.ph..9254C,2009ApJ...698..895K}).

In \cite{2009ApJ...698..895K}, authors analyze $\sim$7.5-year-long MACHO light curves for 15 AGNs and model them with the damped random walk (DRW) model. The resulting AGN variability timescales in that article are comparable to the rest-frame light curve lengths, which likely means they are unreliable (\citealt{2017A&A...597A.128K,2017ApJ...849..110S,2021ApJ...907...96S, Burke_etal_2021Sci}). \cite{2017A&A...597A.128K} showed that time scales derived for the $\sim$9000 SDSS AGNs having eight-year-long light curves are also unreliable.

Because an AGN light curve length is the most important parameter that influences the reliability of the intrinsic time scale measurement (\citealt{2017A&A...597A.128K,2021AcA....71..103K}),
a quest for the longest possible length has begun.
For example \cite{2021ApJ...907...96S}, used the Pan-STARRS1 data to extend the SDSS Stripe 82 quasar light curves to 15 years. A similar approach was used in \cite{Burke_etal_2021Sci}, where the authors used 20-year-long photometric light curves for SDSS Stripe 82 quasars.

The OGLE survey has surveyed the sky since 1992, and the Magellanic Clouds since 1997. There exist AGN light curves from OGLE spanning 26 years and are continuously growing. By providing the physical parameters for these sources, this article sets a pathway to the forthcoming studies on the relations between the physical AGN parameters and variability parameters.

\section{Summary} 
\label{sec:summary}

In this paper, we reanalyzed $\sim$4000 spectra from the Magellanic Quasars Survey. In addition to the already reported 758 AGNs in \cite{2013ApJ...775...92K}, we discovered 161 new AGNs, albeit very faint, so the total number of the MQS AGNs increases to 919.

The spectra for these 919 AGNs were fit with the {\sc PYQSOFIT} code to measure the FWHM (and EW) for the broad lines common in AGNs: \ha{}, \hb{}, \mg{}, and \civ{}, but also EW of FeII blend (reported as ratio to the EW \hb{}; $R_{\rm FeII}$).

Since the spectra were flux-uncalibrated (by design), we used empirical conversions of the broad-band extinction-corrected $V-$ and $I-$band mean OGLE magnitudes to the monochromatic luminosities from \cite{2015AcA....65..251K}. For all the sources, we also calculated the bolometric luminosities, $k-$corrections, distance moduli, and absolute magnitudes.

By combining the broad line FWHM with the monochromatic luminosities, we calculated the black hole masses for 165 AGNs, where the spectra had adequate quality to do so. Whenever two black hole mass measurements were simultaneously available from a single spectrum (\ha{}--\hb{}, \hb{}--\mg{}, or \mg{}--\civ{}), we checked if the two masses stayed in agreement, which was the case.

We also demonstrate the optical plane of the Eigenvector 1, or the quasar main sequence for the sub-sample (41/165) of our sources with reliable measurements of the FWHM(\hb{}) and the strength of the optical \feii{} emission, i.e., \rfe{}. There is an overall agreement with the SDSS-based main sequence diagram where we notice a discernible trend -- increasing Eddington ratio with an increase/decrease in the \rfe{}/FWHM(\hb{}), along the main sequence as found in earlier works \citep{2015ApJ...804L..15S, mar18, panda19b, panda19c, 2023arXiv231003544Z}. The sources with \rfe{} $\gtrsim$ 1 will be especially interesting to follow-up in the optical and near-infrared spectral regions to characterize their variable nature and evaluate the strength of other low-ionization lines, e.g., Ca{\sc ii} triplet (emitting at $\lambda$8498, $\lambda$8542, and $\lambda$8662) and O{\sc i}$\lambda$8446 which are efficient proxy to reveal the physical conditions of the low-ionization line emitting region in such AGNs \citep{2015ApJS..217....3M, 2016ApJ...820..116M, 2020MNRAS.494.4187M, 2021A&A...650A.154P, 2021POBeo.100..287M, 2021POBeo.100..333P, 2021ApJ...918...29M}. Such targets have also been found to be of use to standardize the BLR radius-luminosity relation that can allow to employ quasars as standardizable distance indicators \citep{2019ApJ...883..170M, 2019ApJ...886...42D, 2023FrASS..1030103P, 2023arXiv231113298P}.

The physical parameters of those AGNs, such as the black hole mass, the Eddington luminosity, or the bolometric luminosity will be invaluable for future AGN variability studies. The OGLE survey alone has collected for some of these sources 26-years-long light curves (up to 19-years-long rest-frame) in  $I-$band and slightly shorter $V-$band light curves which will be demonstrated in a forthcoming work.

\begin{acknowledgments}
SP acknowledges the financial support of the Conselho Nacional de Desenvolvimento Científico e Tecnológico (CNPq) Fellowships {300936/2023-0} and {301628/2024-6}. SK has been supported by the Polish National Science Centre through grant number 2018/31/B/ST9/00334 and by the IDUB ``Nowe Idee 2B'' grant from the University of Warsaw, Poland. 
\end{acknowledgments}

\vspace{5mm}
\facilities{Anglo Australian Telescope, 1.3m Warsaw Telescope}

\software{{\sc MATPLOTLIB}  (\citealt{hunter07}); {\sc NUMPY} (\citealt{numpy}); {\sc SCIPY} (\citealt{scipy}); {\sc ASTROPY} (\citealt{astropy}); {\sc TOPCAT} (\citealt{topcat}); {\sc PYQSOFIT} (\citealt{pyqsofit})}

\bibliography{MQS_BHM}{}
\bibliographystyle{aasjournal}

\appendix

%%%%%%%%%%% table 0 - new 161 AGNs
\input{table0_161_new_lim}
\clearpage
%%%%%%%%%%% table 1 - observational parameters
\input{table1_new_lim}
%\clearpage
%%%%%%%%%%% table 2 - PyQSOFit fit parameters
\input{table2_new_lim}
%\clearpage
%%%%%%%%%%% table 3 - Derived parameters Hbeta
\input{table3_new2_lim}

\end{document}

%% file: table_lum_vs_z.tex
% \usepackage{graphicx}
\begin{table}[]
\centering
\caption{Properties of the monochromatic luminosity and redshift distributions}
\label{tab:table_lum_vs_z}
\resizebox{\columnwidth}{!}{%
\begin{tabular}{lcccr}
      & N   & Median & 16$^{\rm th}$ percentile & 84$^{\rm th}$ percentile \\ \hline
log L$_{5100}$ & 78  & {44.120} & {43.720}          & {44.694}          \\
z     & 78  & 0.370  & 0.220           & 0.611           \\ \hline
log L$_{3000}$ & 117 & {45.358} & {44.705}          & {45.789}          \\
z     & 117 & 1.172  & 0.607           & 1.544           \\ \hline
log L$_{1350}$ & 25  & {45.918} & {45.616}          & {46.293}          \\
z     & 25  & 1.645  & 1.563           & 1.870           \\ \hline
\end{tabular}%
}
\tablecomments{The median, 16th and 84th percentile values are represented in respective panels in Figure \ref{fig:L_v_z} and are truncated to 3 decimal digits.}
\end{table}

%% file: table_SDSS.tex
% \usepackage{graphicx}
\begin{table}[]
\centering
\caption{Properties from the SDSS DR14 QSO sample (original versus filtered sample)}
\label{tab:table_SDSS}
\resizebox{\columnwidth}{!}{%
\begin{tabular}{lccr}
original (526,265) & median & min    & max    \\ \hline
z                  & 1.837  & 0.004  & 6.968  \\
log $L_{\rm bol}$               & 45.938 & 29.745 & 48.304 \\
log $M_{\rm BH}$                & 8.684  & 1.089  & 11.459 \\ \hline\\
filtered (449,863) & median & min    & max    \\ \hline
z                  & 1.784  & 0.038  & 5.033  \\
log $L_{\rm bol}$               & 45.954 & 42.979 & 48.304 \\
log $M_{\rm BH}$                & 8.675  & 5.991  & 11.050 \\ \hline
\end{tabular}%
}
\tablecomments{The median, minimum, and maximum values are truncated to 3 decimal digits. The filtered sample is prepared using the {\sc qual} flag = 0 for the \lbol{} and \mbh{} simultaneously. The numbers in the parentheses denote the sources in each sample.}
\end{table}

%% file: table_Eddr_MBH_MQS166.tex
% \usepackage{graphicx}
\begin{table}[]
\centering
\caption{Properties from the \ledd{}-\mbh{} distribution for our sample}
\label{tab:table_Eddr_MBH}
\resizebox{\columnwidth}{!}{%
\begin{tabular}{lccr}
\hb{}, 5100\AA\ (70) & median & min    & max    \\ \hline
log \ledd{}       & $-${1.220} & $-${2.859} & $-${0.177} \\
log \mbh{}        & {8.234}  & {7.271}  & {9.515}  \\ \hline\\
\mg{}, 3000\AA\ (97)   & median & min    & max    \\ \hline
log \ledd{}      & $-${0.795} & $-${1.615} & $-${0.106} \\
log \mbh{}       & {8.838}  & {8.051}  & {9.495}  \\ \hline\\
\civ{}, 1350\AA\ (7)   & median & min    & max    \\ \hline
log \ledd{}      & $-${0.124} & $-${0.721} & {0.130}  \\
log \mbh{}       & {8.778}  & {8.523}  & {9.295}  \\ \hline
\end{tabular}%
}
\tablecomments{The median, minimum, and maximum values are truncated to 3 decimal digits. The numbers in the parentheses denote the sources in each sub-sample.}
\end{table}

%% file: table_QMS.tex
% \usepackage{graphicx}
\begin{table}[]
\centering
\caption{Properties of the sample presented in the Quasar Main Sequence diagram in Figure \ref{fig:EV1}}
\label{tab:table_qms}
\resizebox{\columnwidth}{!}{%
\begin{tabular}{lccr}
MQS (41)     & median   & min       & max       \\ \hline
z            & 0.351    & 0.151     & 0.647     \\
log $L_{\rm 5100}$    & {44.090}   & {43.453}    & {45.318}    \\
FWHM(\hb{})     & 3784.782 & 1436.520  & 11526.192 \\
log \mbh{}      & {8.331}    & {7.273}     & {9.205}     \\
log \ledd{}     & $-${1.276}   & $-${2.367}    & $-${0.414}    \\
\rfe{}         & 0.526    & $<$ 0.001 & 2.378     \\ \hline\\
%6.809e-08
SDSS (18762) & median   & min       & max       \\ \hline
z            & 0.581    & 0.056     & 0.890     \\
log $L_{\rm 5100}$    & 44.539   & 42.769    & 46.267    \\
FWHM(\hb{})     & 3709.279 & 925.010   & 18263.505 \\
log \mbh{}       & 8.341    & 6.560     & 9.936     \\
log \ledd{}     & $-$0.952   & $-$2.757    & 0.451     \\
\rfe{}          & 0.713    & 0.037     & 6.142     \\ \hline
\end{tabular}%
}
\tablecomments{The median, minimum, and maximum values are truncated to 3 decimal digits. The numbers in the parenthesis denote the sources in each sample.}
\end{table}

%% file: table0_161_new_lim.tex
\begin{sidewaystable}
    %\centering
\startlongtable
\begin{deluxetable}{lccccccccccccccr}
%% This is the title of the table.
\tabletypesize{\tiny}
\tablecaption{Observational parameters for the new 161 MQS AGNs (full version available electronically).}
\tablenum{A1}
\label{table0-obs}

\tablehead{\colhead{Name} & \colhead{RA} & \colhead{Dec} & \colhead{z} & \colhead{DM} & \colhead{mean V} & \colhead{mean I} & \colhead{Ext V} & \colhead{Ext I} & \colhead{K corr V} & \colhead{K corr I} & \colhead{Mag V} & \colhead{Mag I} & \colhead{L$_{5100\AA}$} & \colhead{L$_{3000\AA}$} & \colhead{L$_{1350\AA}$}\\ 
\colhead{} & \colhead{hh:mm:ss.ss} & \colhead{dd:mm:ss.ss} & \colhead{} & \colhead{Mpc} & \colhead{(obs)} & \colhead{(obs)} & \colhead{} & \colhead{} & \colhead{} & \colhead{} & \colhead{(abs.)} & \colhead{(abs.)} & \colhead{(erg s$^{-1}$)} & \colhead{(erg s$^{-1}$)} & \colhead{(erg s$^{-1}$)}} 

%% All data must appear between the \startdata and \enddata commands
\startdata
MQS J043443.27$-$695400.0 & 04:34:43.27 & $-$69:54:00.00 & 1.215 & 44.653 & 19.979 & 19.414 & 0.235 & 0.129 & $-$0.395 & 0.087 & $-${24.749} & $-${25.584} & $-$$-$ & {45.346}$\pm$0.051 & $-$$-$ \\
MQS J043445.71$-$690859.9 & 04:34:45.71 & $-$69:08:59.90 & 0.562 & 42.584 & 21.83 & 20.635 & 0.206 & 0.113 & $-$0.199 & 0.185 & $-${20.967} & $-${22.36} & {43.772}$\pm$0.073 & {43.95}$\pm$0.07 & $-$$-$ \\
MQS J043501.08$-$701323.8 & 04:35:01.08 & $-$70:13:23.80 & 1.335 & 44.908 & $-$$-$ & $-$$-$ & 0.259 & 0.142 & $-$0.385 & 0.017 & $-$$-$ & $-$$-$ & $-$$-$ & $-$$-$ & $-$$-$ \\
MQS J043522.52$-$703223.1 & 04:35:22.52 & $-$70:32:23.10 & 0.412 & 41.773 & 21.717 & 20.295 & 0.364 & 0.2 & $-$0.055 & 0.235 & $-${20.729} & $-${22.113} & {43.704}$\pm$0.075 & {43.831}$\pm$0.091 & $-$$-$ \\
MQS J043541.79$-$702931.7 & 04:35:41.79 & $-$70:29:31.70 & 1.345 & 44.928 & 20.197 & 19.251 & 0.337 & 0.185 & $-$0.385 & 0.013 & $-${25.02} & $-${26.06} & $-$$-$ & {45.493}$\pm$0.044 & $-$$-$ \\
MQS J043553.04$-$685953.7 & 04:35:53.04 & $-$68:59:53.70 & 0.723 & 43.253 & $-$$-$ & $-$$-$ & 0.251 & 0.138 & $-$0.29 & 0.167 & $-$$-$ & $-$$-$ & $-$$-$ & $-$$-$ & $-$$-$ \\
MQS J043617.19$-$701031.8 & 04:36:17.19 & $-$70:10:31.80 & 0.967 & 44.036 & 19.141 & 18.501 & 0.284 & 0.156 & $-$0.423 & 0.223 & $-${25.04} & $-${26.07} & $-$$-$ & {45.5}$\pm$0.048 & $-$$-$ \\
MQS J043625.99$-$700735.9 & 04:36:25.99 & $-$70:07:35.90 & 1.777 & 45.678 & 20.734 & 18.722 & 0.284 & 0.156 & $-$0.512 & $-$0.162 & $-${25} & $-${27.106} & $-$$-$ & {45.741}$\pm$0.053 & {45.874}$\pm$0.082 \\
MQS J043635.15$-$693656.5 & 04:36:35.15 & $-$69:36:56.50 & 1.297 & 44.83 & $-$$-$ & $-$$-$ & 0.251 & 0.138 & $-$0.388 & 0.035 & $-$$-$ & $-$$-$ & $-$$-$ & $-$$-$ & $-$$-$ \\
MQS J043657.20$-$703824.7 & 04:36:57.20 & $-$70:38:24.70 & 0.657 & 42.998 & 21.416 & 20.845 & 0.397 & 0.218 & $-$0.251 & 0.173 & $-${22.125} & $-${22.762} & {44.058}$\pm$0.074 & {44.269}$\pm$0.058 & $-$$-$ \\
MQS J043742.14$-$693310.1 & 04:37:42.14 & $-$69:33:10.10 & 0.97 & 44.044 & 19.803 & 19.502 & 0.244 & 0.134 & $-$0.424 & 0.222 & $-${24.305} & $-${25.032} & $-$$-$ & {45.153}$\pm$0.048 & $-$$-$ \\
MQS J043825.54$-$703614.6 & 04:38:25.54 & $-$70:36:14.60 & 1.49 & 45.204 & 20.234 & 19.151 & 0.41 & 0.225 & $-$0.385 & $-$0.039 & $-${25.405} & $-${26.464} & $-$$-$ & {45.644}$\pm$0.049 & $-$$-$ \\
MQS J043845.15$-$692821.1 & 04:38:45.15 & $-$69:28:21.10 & 1.361 & 44.96 & 20.903 & 20.071 & 0.295 & 0.162 & $-$0.386 & 0.008 & $-${24.261} & $-${25.221} & $-$$-$ & {45.174}$\pm$0.05 & $-$$-$ \\
MQS J043900.74$-$693906.6 & 04:39:00.74 & $-$69:39:06.60 & 1.306 & 44.848 & 19.749 & 19.047 & 0.302 & 0.166 & $-$0.388 & 0.03 & $-${25.315} & $-${26.163} & $-$$-$ & {45.576}$\pm$0.051 & $-$$-$ \\
MQS J043904.58$-$680349.6 & 04:39:04.58 & $-$68:03:49.60 & 0.846 & 43.675 & $-$$-$ & 19.436 & 0.164 & 0.09 & $-$0.376 & 0.24 & $-$$-$ & $-${24.659} & $-$$-$ & {44.61}$\pm$-- & $-$$-$ \\
MQS J043908.00$-$683015.0 & 04:39:08.00 & $-$68:30:15.00 & 1.358 & 44.954 & 21.381 & 20.701 & 0.191 & 0.105 & $-$0.385 & 0.009 & $-${23.57} & $-${24.472} & $-$$-$ & {44.886}$\pm$0.05 & $-$$-$ \\
MQS J043925.47$-$674256.1 & 04:39:25.47 & $-$67:42:56.10 & 1.42 & 45.074 & 20.077 & 18.69 & 0.177 & 0.097 & $-$0.382 & $-$0.01 & $-${24.969} & $-${26.568} & $-$$-$ & {45.595}$\pm$0.046 & $-$$-$ \\
MQS J044002.72$-$681538.4 & 04:40:02.72 & $-$68:15:38.40 & 0.557 & 42.56 & 21.679 & 20.452 & 0.155 & 0.085 & $-$0.195 & 0.187 & $-${20.996} & $-${22.465} & {43.802}$\pm$0.073 & {43.979}$\pm$0.07 & $-$$-$ \\
MQS J044003.65$-$673216.0 & 04:40:03.65 & $-$67:32:16.00 & 0.759 & 43.384 & $-$$-$ & $-$$-$ & 0.151 & 0.083 & $-$0.336 & 0.186 & $-$$-$ & $-$$-$ & $-$$-$ & $-$$-$ & $-$$-$ \\
MQS J044055.81$-$681833.7 & 04:40:55.81 & $-$68:18:33.70 & 0.409 & 41.754 & 21.374 & 20.142 & 0.155 & 0.085 & $-$0.052 & 0.237 & $-${20.638} & $-${22.019} & {43.667}$\pm$0.075 & {43.793}$\pm$0.091 & $-$$-$ \\
MQS J044142.86$-$673239.6 & 04:41:42.86 & $-$67:32:39.60 & 0.972 & 44.05 & $-$$-$ & 20.706 & 0.137 & 0.075 & $-$0.425 & 0.221 & $-$$-$ & $-${23.715} & $-$$-$ & {44.235}$\pm$-- & $-$$-$ \\
MQS J044153.90$-$684535.3 & 04:41:53.90 & $-$68:45:35.30 & 1.78 & 45.683 & 20.503 & 19.788 & 0.255 & 0.14 & $-$0.513 & $-$0.162 & $-${25.177} & $-${26.013} & $-$$-$ & {45.513}$\pm$0.053 & {45.65}$\pm$0.082 \\
MQS J044245.58$-$693000.6 & 04:42:45.58 & $-$69:30:00.60 & 1.199 & 44.617 & 20.049 & 19.364 & 0.324 & 0.178 & $-$0.395 & 0.097 & $-${24.821} & $-${25.706} & $-$$-$ & {45.383}$\pm$0.05 & $-$$-$ \\
MQS J044312.02$-$694537.4 & 04:43:12.02 & $-$69:45:37.40 & 1.61 & 45.413 & 19.965 & 19.115 & 0.282 & 0.155 & $-$0.449 & $-$0.114 & $-${25.563} & $-${26.494} & $-$$-$ & {45.69}$\pm$0.047 & {45.838}$\pm$0.092 \\
MQS J044416.36$-$684215.0 & 04:44:16.36 & $-$68:42:15.00 & 2.54 & 46.631 & 20.318 & 19.522 & 0.321 & 0.176 & $-$0.696 & $-$0.162 & $-${26.259} & $-${27.299} & $-$$-$ & $-$$-$ & {46.088}$\pm$0.06 \\
MQS J044417.49$-$693700.4 & 04:44:17.49 & $-$69:37:00.40 & 1.429 & 45.091 & 19.726 & 18.979 & 0.288 & 0.158 & $-$0.382 & $-$0.013 & $-${25.559} & $-${26.415} & $-$$-$ & {45.671}$\pm$0.049 & $-$$-$ \\
MQS J044515.85$-$672535.2 & 04:45:15.85 & $-$67:25:35.20 & 0.555 & 42.551 & 20.261 & 20.692 & 0.211 & 0.116 & $-$0.194 & 0.19 & $-${22.518} & $-${22.281} & {44.144}$\pm$0.073 & {44.319}$\pm$0.07 & $-$$-$ \\
MQS J044548.15$-$694041.9 & 04:45:48.15 & $-$69:40:41.90 & 0.35 & 41.355 & 20.337 & 18.984 & 0.361 & 0.198 & $-$0.007 & 0.193 & $-${21.733} & $-${22.96} & {44.1}$\pm$0.091 & $-$$-$ & $-$$-$ \\
MQS J044600.79$-$681704.8 & 04:46:00.79 & $-$68:17:04.80 & 0.549 & 42.522 & 21.238 & 20.782 & 0.288 & 0.158 & $-$0.19 & 0.185 & $-${21.67} & $-${22.241} & {43.903}$\pm$0.065 & {44.076}$\pm$0.067 & $-$$-$ \\
MQS J044644.59$-$673544.4 & 04:46:44.59 & $-$67:35:44.40 & 1.354 & 44.946 & 20.849 & 19.902 & 0.235 & 0.129 & $-$0.385 & 0.01 & $-${24.182} & $-${25.312} & $-$$-$ & {45.176}$\pm$0.047 & $-$$-$ \\
MQS J044726.40$-$673055.6 & 04:47:26.40 & $-$67:30:55.60 & 1.667 & 45.507 & 20.34 & 19.297 & 0.259 & 0.142 & $-$0.479 & $-$0.144 & $-${25.206} & $-${26.35} & $-$$-$ & {45.592}$\pm$0.049 & {45.736}$\pm$0.087 \\
MQS J044758.66$-$664953.0 & 04:47:58.66 & $-$66:49:53.00 & 0.78 & 43.457 & $-$$-$ & $-$$-$ & 0.259 & 0.142 & $-$0.355 & 0.198 & $-$$-$ & $-$$-$ & $-$$-$ & $-$$-$ & $-$$-$ \\
MQS J044758.91$-$671242.1 & 04:47:58.91 & $-$67:12:42.10 & 1.525 & 45.267 & 21.265 & 20.148 & 0.295 & 0.162 & $-$0.39 & $-$0.052 & $-${24.202} & $-${25.391} & $-$$-$ & {45.187}$\pm$0.053 & {45.381}$\pm$0.073 \\
MQS J044827.18$-$694337.7 & 04:48:27.18 & $-$69:43:37.70 & 0.256 & 40.569 & 21.702 & 19.329 & 0.308 & 0.169 & 0.013 & $-$0.015 & $-${19.496} & $-${21.563} & {43.455}$\pm$0.086 & $-$$-$ & $-$$-$ \\
MQS J044830.07$-$663837.4 & 04:48:30.07 & $-$66:38:37.40 & 0.449 & 41.996 & 20.085 & 19.266 & 0.184 & 0.101 & $-$0.092 & 0.238 & $-${22.187} & $-${23.17} & {44.196}$\pm$0.07 & {44.334}$\pm$0.082 & $-$$-$ \\
MQS J044922.60$-$664435.3 & 04:49:22.60 & $-$66:44:35.30 & 0.595 & 42.735 & 21.807 & 20.188 & 0.186 & 0.102 & $-$0.217 & 0.178 & $-${21.083} & $-${22.929} & {43.921}$\pm$0.067 & {44.11}$\pm$0.069 & $-$$-$ \\
MQS J045051.09$-$671505.8 & 04:50:51.09 & $-$67:15:05.80 & 1.645 & 45.471 & 20.279 & 19.357 & 0.204 & 0.112 & $-$0.469 & $-$0.136 & $-${25.131} & $-${26.202} & $-$$-$ & {45.547}$\pm$0.052 & {45.693}$\pm$0.091 \\
MQS J045058.16$-$671634.4 & 04:50:58.16 & $-$67:16:34.40 & 0.979 & 44.069 & $-$$-$ & $-$$-$ & 0.204 & 0.112 & $-$0.427 & 0.218 & $-$$-$ & $-$$-$ & $-$$-$ & {45.538}$\pm$0.052 & $-$$-$ \\
MQS J045107.05$-$673800.9 & 04:51:07.05 & $-$67:38:00.90 & 1.665 & 45.503 & 21.039 & 20.097 & 0.191 & 0.105 & $-$0.479 & $-$0.143 & $-${24.367} & $-${25.473} & $-$$-$ & {45.249}$\pm$0.049 & {45.393}$\pm$0.087 \\
MQS J045242.83$-$674236.2 & 04:52:42.83 & $-$67:42:36.20 & 0.492 & 42.234 & 21.823 & 20.576 & 0.12 & 0.066 & $-$0.136 & 0.211 & $-${20.515} & $-${22.001} & {43.626}$\pm$0.073 & {43.78}$\pm$0.074 & $-$$-$ \\
MQS J045253.95$-$674504.9 & 04:52:53.95 & $-$67:45:04.90 & 1.334 & 44.906 & $-$$-$ & $-$$-$ & 0.126 & 0.069 & $-$0.386 & 0.017 & $-$$-$ & $-$$-$ & $-$$-$ & $-$$-$ & $-$$-$ \\
MQS J045424.28$-$680952.7 & 04:54:24.28 & $-$68:09:52.70 & 0.319 & 41.12 & 21.789 & 21.18 & 0.284 & 0.156 & 0.003 & 0.049 & $-${19.902} & $-${20.301} & {43.254}$\pm$0.077 & $-$$-$ & $-$$-$ \\
MQS J045525.06$-$704421.6 & 04:55:25.06 & $-$70:44:21.60 & 0.769 & 43.419 & 20.222 & 19.504 & 0.268 & 0.147 & $-$0.347 & 0.192 & $-${23.386} & $-${24.401} & {44.593}$\pm$0.075 & {44.843}$\pm$0.055 & $-$$-$ \\
MQS J045632.58$-$670100.8 & 04:56:32.58 & $-$67:01:00.80 & 1.087 & 44.352 & 20.144 & 19.6 & 0.224 & 0.123 & $-$0.426 & 0.175 & $-${24.23} & $-${25.173} & $-$$-$ & {45.159}$\pm$0.046 & $-$$-$ \\
MQS J045651.08$-$680918.0 & 04:56:51.08 & $-$68:09:18.00 & 1.34 & 44.918 & 20.897 & 20.221 & 0.293 & 0.161 & $-$0.385 & 0.015 & $-${24.222} & $-${25.034} & $-$$-$ & {45.131}$\pm$0.044 & $-$$-$ \\ \enddata

\vspace{0.5cm}

%% General table comment marker
\tablecomments{Columns are as follows: (1) MQS name; (2) right ascension (RA) in hh:mm:ss.ss; (3) declination (Dec) in dd:mm:ss.ss; (4) redshift (z); (5) distance modulus (DM); (6) mean observed magnitude in V$-$band; (7) mean observed magnitude in I$-$band; (8) Extinction in V$-$band; (9) Extinction in I$-$band; (10) K$-$correction in V$-$band; (11) K$-$correction in I$-$band; (12) absolute magnitude in V$-$band; (13) absolute magnitude in I$-$band; (14) monochromatic AGN luminosity estimated at 5100\AA~using photometric scaling (in log$-$scale); (15) monochromatic AGN luminosity estimated at 3000\AA~using photometric scaling (in log$-$scale); and (16) monochromatic AGN luminosity estimated at 1350\AA~using photometric scaling (in log$-$scale).}

%% General table references marker
%\tablerefs{table1$-$obs}

\end{deluxetable}
\end{sidewaystable}

%% file: table1_new_lim.tex
\startlongtable
\begin{deluxetable}{lcccccccccccr}

%% This is the title of the table.
\tabletypesize{\tiny}
\tablecaption{Observational parameters for the 165 MQS AGNs (full version available electronically).}
\tablenum{A2}
\label{table1-obs}

\tablehead{\colhead{Name} & \colhead{RA} & \colhead{Dec} & \colhead{z} & \colhead{DM} & \colhead{mean V} & \colhead{mean I} & \colhead{Ext V} & \colhead{Ext I} & \colhead{K corr V} & \colhead{K corr I} & \colhead{Mag V} & \colhead{Mag I}\\ 
\colhead{} & \colhead{hh:mm:ss.ss} & \colhead{dd:mm:ss.ss} & \colhead{} & \colhead{Mpc} & \colhead{(obs)} & \colhead{(obs)} & \colhead{} & \colhead{} & \colhead{} & \colhead{} & \colhead{(abs.)} & \colhead{(abs.)}} 

%% All data must appear between the \startdata and \enddata commands
\startdata
%MQS\_J053242.46$-$692612.2 & 05:32:42.46 & $-$69:26:12.20 & 0.059 & 37.11  & 21.177  & 17.47   & 0.222 & 0.122 & $-$0.032 & 0.064  & $-$16.123 & $-$19.826 \\
MQS J053159.69$-$691951.6 & 05:31:59.69 & $-$69:19:51.60 & 0.149 & 39.255 & 18.96   & 18.18   & 0.228 & 0.125 & $-$0.065 & $-$0.062 & $-${20.686} & $-${21.263} \\
MQS J052402.28$-$701108.7 & 05:24:02.28 & $-$70:11:08.70 & 0.151 & 39.287 & 17.942  & 17.193  & 0.18  & 0.099 & $-$0.062 & $-$0.062 & $-${21.643} & $-${22.230} \\
MQS J050502.09$-$694504.0 & 05:05:02.09 & $-$69:45:04.00 & 0.155 & 39.349 & 20.151  & 18.659  & 0.268 & 0.147 & $-$0.06  & $-$0.061 & $-${19.674} & $-${20.923} \\
MQS J050634.04$-$691048.3 & 05:06:34.04 & $-$69:10:48.30 & 0.16  & 39.425 & 21.242  & 17.62   & 0.228 & 0.125 & $-$0.059 & $-$0.056 & $-${18.580} & $-${21.999} \\ 
MQS J051716.95$-$704402.0 & 05:17:16.95 & $-$70:44:02.00 & 0.169 & 39.556 & 19.626  & 17.77   & 0.191 & 0.105 & $-$0.055 & $-$0.055 & $-${20.257} & $-${21.941} \\ \enddata
%MQS\_J005551.23$-$733110.3 & 00:55:51.23 & $-$73:31:10.30 & 0.186 & 39.787 & 21.517  & 18.05   & 0.102 & 0.056 & $-$0.036 & $-$0.045 & $-$18.336 & $-$21.748 \\
%MQS\_J045538.57$-$690455.1 & 04:55:38.57 & $-$69:04:55.10 & 0.189 & 39.825 & 18.985  & 18.144  & 0.315 & 0.173 & $-$0.032 & $-$0.041 & $-$21.123 & $-$21.813 \\
%MQS\_J054917.21$-$713049.4 & 05:49:17.21 & $-$71:30:49.40 & 0.19  & 39.838 & 20.706  & 18.958  & 0.308 & 0.169 & $-$0.03  & $-$0.041 & $-$19.41  & $-$21.008 \\
%MQS\_J011139.65$-$725031.9 & 01:11:39.65 & $-$72:50:31.90 & 0.198 & 39.938 & 19.797  & 18.659  & 0.168 & 0.092 & $-$0.02  & $-$0.033 & $-$20.289 & $-$21.338 \\ \enddata

%% General table comment marker
\tablecomments{Columns are as follows: (1) MQS name; (2) right ascension (RA) in hh:mm:ss.ss; (3) declination (Dec) in dd:mm:ss.ss; (4) redshift (z); (5) distance modulus (DM); (6) mean observed magnitude in V$-$band; (7) mean observed magnitude in I$-$band; (8) Extinction in V$-$band; (9) Extinction in I$-$band; (10) K$-$correction in V$-$band; (11) K$-$correction in I$-$band; (12) absolute magnitude in V$-$band; and (13) absolute magnitude in I$-$band}

%% General table references marker
%\tablerefs{table1$-$obs}

\end{deluxetable}

%% file: table2_new_lim.tex
\startlongtable
\begin{deluxetable}{lccccccr}

%% This is the title of the table.
\tabletypesize{\tiny}
\tablecaption{AGN monochromatic luminosities and FWHMs for prominent broad emission lines for the 165 MQS AGNs (full version available electronically).}
\tablenum{A3}
\label{table2-obs}

\tablehead{\colhead{Name} & \colhead{L$_{5100\AA}$} & \colhead{L$_{3000\AA}$} & \colhead{L$_{1350\AA}$} & \colhead{FWHM(H$\alpha$)} & \colhead{FWHM(H$\beta$)} & \colhead{FWHM(MgII)} & \colhead{FWHM(CIV)}\\ 
\colhead{} & \colhead{(erg s$^{-1}$)} & \colhead{(erg s$^{-1}$)} & \colhead{(erg s$^{-1}$)} & \colhead{(km s$^{-1}$)} & \colhead{(km s$^{-1}$)} & \colhead{(km s$^{-1}$)} & \colhead{(km s$^{-1}$)}} 

%% All data must appear between the \startdata and \enddata commands
\startdata
MQS J053159.69$-$691951.6 & {43.587}$\pm$0.073 & -- & -- & 2759$\pm$10 & 3373$\pm$30 & -- & -- \\
MQS J052402.28$-$701108.7 & {43.971}$\pm$0.073 & -- & -- & 7617$\pm$1382 & 11526$\pm$64 & -- & -- \\
MQS J050502.09$-$694504.0 & {43.310}$\pm$0.073 & -- & -- & -- & 3386$\pm$4915 & -- & -- \\
MQS J050634.04$-$691048.3 & {43.527}$\pm$0.059 & -- & -- & 9347$\pm$767 & 3919$\pm$49 & -- & -- \\
MQS J051716.95$-$704402.0 & {43.649}$\pm$0.076 & -- & -- & 2831$\pm$27 & 4376$\pm$703 & $-$- & -- \\ \enddata

%% General table comment marker
\tablecomments{Columns are as follows: (1) MQS name; (2) monochromatic AGN luminosity estimated at 5100\AA~using photometric scaling (in log-scale); (3) monochromatic AGN luminosity estimated at 3000\AA~using photometric scaling (in log-scale); (4) monochromatic AGN luminosity estimated at 1350\AA~using photometric scaling (in log-scale); (5) Full-width at half maximum (FWHM) of the broad H$\alpha$ profile; (6) FWHM of the broad H$\beta$ profile; (7) FWHM of the broad MgII profile; and (8) FWHM of the broad CIV profile.}

%% General table references marker
%\tablerefs{table1-obs}

\end{deluxetable}

%% file: table3_new2_lim.tex
\startlongtable
\begin{deluxetable}{lcccccccccr}
\addtolength{\tabcolsep}{-0.25em}
%% This is the title of the table.
\tabletypesize{\tiny}
\tablecaption{AGN bolometric luminosities, black hole masses, and Eddington ratios for the 165 MQS AGNs. (full version available electronically).}
\tablenum{A4}
\label{table3-obs}

\tablehead{\colhead{Name} & \colhead{L$_{\rm bol}^{5100\AA}$} & \colhead{L$_{\rm bol}^{3000\AA}$} & \colhead{L$_{\rm bol}^{1350\AA}$} & \colhead{M$_{\rm BH}^{\rm H\alpha}$} & \colhead{M$_{\rm BH}^{\rm H\beta}$} & \colhead{M$_{\rm BH}^{\rm MgII}$} & \colhead{M$_{\rm BH}^{\rm CIV}$} & \colhead{$\lambda^{\rm H\beta}_{\rm Edd}$} & \colhead{$\lambda^{\rm MgII}_{\rm Edd}$} & \colhead{$\lambda^{\rm CIV}_{\rm Edd}$}\\ 
\colhead{} & \colhead{(erg s$^{-1}$)} & \colhead{(erg s$^{-1}$)} & \colhead{(erg s$^{-1}$)} & \colhead{(M$_{\odot}$)} & \colhead{(M$_{\odot}$)} & \colhead{(M$_{\odot}$)} & \colhead{(M$_{\odot}$)} & \colhead{(5100\AA)} & \colhead{(3000\AA)} & \colhead{(1350\AA)}} 

%% All data must appear between the \startdata and \enddata commands
\startdata
MQS J053159.69$-$691951.6 & {44.554}$\pm$0.073 & -- & -- & {7.709}$\pm$0.040 & {7.892}$\pm$0.044 & -- & -- & $-${1.439}$\pm$0.117 & -- & -- \\
MQS J052402.28$-$701108.7 & {44.938}$\pm$0.073 & -- & -- & {8.827}$\pm$0.194 & {9.205}$\pm$0.041 & -- & -- & $-${2.367}$\pm$0.114 & -- & -- \\
MQS J050502.09$-$694504.0 & {44.277}$\pm$0.073 & -- & -- & -- & {7.757}$\pm$1.297 & -- & -- & $-${1.581}$\pm$1.369 & -- & -- \\
MQS J050634.04$-$691048.3 & {44.494}$\pm$0.059 & -- & -- & {8.792}$\pm$0.101 & {7.999}$\pm$0.040 & -- & -- & $-${1.606}$\pm$0.099 & -- & -- \\
MQS J051716.95$-$704402.0 & {44.616}$\pm$0.076 & -- & -- & {7.764}$\pm$0.046 & {8.161}$\pm$0.177 & -- & -- & $-${1.645}$\pm$0.253 & -- & -- \\ \enddata
%MQS\_J005551.23-733110.3 & 43.588$\pm$0.005 & -- & -- & -- & 7.250$\pm$0.000 & -- & -- & -1.762$\pm$0.000 & -- & -- \\
%MQS\_J045538.57-690455.1 & 44.749$\pm$0.006 & -- & -- & -- & 7.889$\pm$0.004 & -- & -- & -1.241$\pm$0.000 & -- & -- \\
%MQS\_J054917.21-713049.4 & 44.255$\pm$0.006 & -- & -- & 7.812$\pm$0.001 & -- & -- & -- & -- & -- & -- \\
%MQS\_J011139.65-725031.9 & 44.481$\pm$0.004 & -- & -- & 7.414$\pm$0.017 & -- & -- & -- & -- & -- & -- \\ \enddata
%% General table comment marker
\tablecomments{Columns are as follows: (1) MQS name; (2) Bolometric luminosity (L$_{\rm bol}$) estimated using the L$_{5100\AA}$ (in log-scale); (3) L$_{\rm bol}$ estimated using the L$_{3000\AA}$ (in log-scale); (4) L$_{\rm bol}$ estimated using the L$_{1350\AA}$ (in log-scale); (5) Black hole mass (M$_{\rm BH}$) estimated using the L$_{5100\AA}$ and FWHM(H$\alpha$) (in log-scale); (6) M$_{\rm BH}$ estimated using the L$_{5100\AA}$ and FWHM(H$\beta$) (in log-scale); (7) M$_{\rm BH}$ estimated using the L$_{3000\AA}$ and FWHM(MgII) (in log-scale); (8) M$_{\rm BH}$ estimated using the L$_{1350\AA}$ and FWHM(CIV) (in log-scale); (9) Eddington ratio ($\lambda_{\rm Edd}$) estimated using the L$_{\rm bol}$ from 5100\AA~and H$\beta$-based M$_{\rm BH}$ (in log-scale); (10) $\lambda_{\rm Edd}$ estimated using the L$_{\rm bol}$ from 3000\AA~and MgII-based M$_{\rm BH}$ (in log-scale); and (11) $\lambda_{\rm Edd}$ estimated using the L$_{\rm bol}$ from 1350\AA~and CIV-based M$_{\rm BH}$ (in log-scale). {We note however that we do not account for the error on the constant term (A, see Equation \ref{eq:mbh}) while estimating the uncertainties on the BH masses.}}

%% General table references marker
%\tablerefs{table1-obs}

\end{deluxetable}